\documentclass[aps,prf,longbibliography,onecolumn]{revtex4-2} %showpacs,notitlepage,nofootinbib,longbibliograph

\usepackage[utf8]{inputenc}
\usepackage{hyperref}
\usepackage{epsf,graphicx}
\usepackage{amsmath,gensymb,amssymb}
\usepackage{siunitx}
\usepackage[bottom]{footmisc}
\usepackage{lipsum}
\usepackage{dcolumn}% Align table columns on decimal point
\usepackage{xcolor}
\usepackage{upgreek}

\usepackage{bm}% bold math
\usepackage{color}
\usepackage[type={CC},modifier={by-nc-sa},version={4.0},]{doclicense}

\usepackage{tikz}
\hypersetup{
    colorlinks = true, %colorise les liens
    urlcolor   = blue, %couleur des hyperliens
    citecolor  = blue,
    urlcolor= blue, %couleur des hyperliens
linkcolor= blue, %couleur des liens internes
bookmarksopen=true, %si les signets Acrobat sont crees,
}
\begin{document}

\title{Finite-system size effects in gravity-capillary wave turbulence}

\author{Tanu Singla}
\affiliation{Universit\'e Paris Cit\'e, CNRS, MSC Laboratory, UMR 7057, F-75013 Paris, France}

\author{Jean-Baptiste Gorce}
\affiliation{Universit\'e Paris Cit\'e, CNRS, MSC Laboratory, UMR 7057, F-75013 Paris, France}

\author{Eric Falcon}
\email{eric.falcon@u-paris.fr}
\affiliation{Universit\'e Paris Cit\'e, CNRS, MSC Laboratory, UMR 7057, F-75013 Paris, France}
%\orcid{0000-0001-9640-9895}

\date{\today}

\begin{abstract}
We experimentally investigate the effects of finite-system size on the dynamics of weakly nonlinear random gravity-capillary surface waves. Experiments are conducted in rectangular tanks with varying aspect ratios, in which the fluid surface is perturbed locally and erratically by small, partially submerged magnets. Driven by an oscillating vertical electromagnetic field, these magnets generate a statistically homogeneous and isotropic random wave field. This setup enables us to probe finite-size effects without the dominant influence of global forcing present in horizontally oscillated tanks. Spatiotemporal measurements of the wave field reveal multiple branches in the wave-energy spectrum along the unconfined direction, corresponding to sloshing modes in the confined direction. We show that the spectral properties of these modes can be tuned by varying either the wave steepness or the confinement. Signatures of discrete wave turbulence in the confined direction and mesoscopic continuous wave turbulence in the unconfined direction are observed. As the confinement is gradually relaxed, we further demonstrate a smooth transition from discrete to continuous wave turbulence, consistent with the nonlinear-to-discreteness timescale ratio. Using high-order correlation analysis, we also show that finite-size effects alter wave dynamics by depleting two-dimensional three-wave resonant interactions along the confined direction.
\end{abstract}

\maketitle

\section{Introduction}\label{intro}
Weak-turbulence theory describes dynamical and statistical properties of random weakly nonlinear dispersive waves in various systems~\cite{Zakharov,NazarenkoBook,GaltierBook}. When wave energy cascades across scales due to resonant nonlinear-wave interactions, the out-of-equilibrium stationary solution of the kinetic equation yields a power-law dependence of the wave-energy spectrum on the scale (Kolmogorov-Zakharov spectrum).  Initiated in the 1960s to model the ocean wave spectrum~\cite{Hasselmann}, this theory has since been applied in almost all systems involving waves, such as ocean surface waves, plasma waves, hydroelastic waves, elastic waves on a plate, internal or inertial waves on rotating stratified fluids, and optical waves~\cite{Zakharov,NazarenkoBook,GaltierBook}.  %ranging from plasmas \cite{Sagdeev}, elastic sheets \cite{Gustavo,Deike,Hassaini1}, optical systems \cite{Picozzi} etc. (see for a recent review on experiments~\cite{FalconARFM2022}.

Weak-turbulence theory requires many assumptions, as an infinite spatial domain or weakly nonlinear waves. The first one is usually not achieved numerically or experimentally.  For an infinitely large system ($L\to \infty$), the Fourier modes $k$ are continuous, and both exact resonant and quasiresonant wave interactions are possible. For finite $L$, exact wave interactions are rare since $k$ modes are discrete and few in number. In this case, quasiresonant wave interactions become dominant for strong enough nonlinearity (but still weak), and the energy cascade is still continuous in the Fourier space, as shown numerically~\cite{PushkarevEPJB1999,DyachenkoJETP2003}. However, when the system size $L$ is too small, Fourier modes become highly discrete, and the spacing between adjacent modes ($\Delta k=\pi/L$) can exceed the nonlinear spectral broadening. In this regime, known as discrete wave turbulence, only exact resonances contribute significantly to the dynamics, as quasi-resonant interactions are weakened by frequency gaps~\cite{ZakharovJETP2005,HrabskiPRE2020}. Exact resonant interactions are then generally strongly depleted, leading to nonlocal energy transfers~\cite{KartashovaPRL1994,KartashovaEPL2009,LvovPRE2010}. The regime in which both continuous and discrete wave turbulence can coexist~\cite{NazarenkoNJP2007,ZhangPRE2022} is termed ``mesoscopic" wave turbulence~\cite{ZakharovJETP2005}. For instance, a bursty transfer of energy across scales, reminiscent of sandpile-like avalanches, has been proposed~\cite{NazarenkoJSM2006}. In the limit of strong discreteness, the system no longer supports a cascade and is thus referred to as ``frozen turbulence''~\cite{PushkarevEPJB1999,PushkarevPhysicaD2000}. Taking these finite-system size effects into account is also an important challenge in pure mathematics~\cite{Faou2016,BanksPRL2022}.

%If $L$ is the system size, wavenumbers $k$ in the Fourier space are thus discrete (i.e., adjacent eigenmodes are spatially separated by $\Delta k=\pi/L$), whereas for an infinitely large system ($L\to \infty$) all Fourier modes are continuous and resonant wave interactions exist. For finite $L$, quasiresonant wave interactions become dominant when nonlinearity is increased (but still weak), and the kinetic regime can be recovered, leading to a continuous energy cascade in the Fourier space, as shown numerically~\cite{PushkarevEPJB1999,DyachenkoJETP2003}. If $L$ becomes too small, the spacing between discrete Fourier modes becomes larger than the nonlinear broadening. The system thus enters the regime of discrete wave turbulence, and only exact resonant interactions contribute significantly to the dynamics, as quasiresonant interactions are suppressed by the frequency gaps~\cite{ZakharovJETP2005,HrabskiPRE2020}. 

%If $L$ is not large enough for a fixed nonlinearity level, the discreteness of the Fourier modes becomes dominant over the nonlinear broadening, giving rise to discrete wave turbulence where only exact resonances are involved (as quasiresonances are prohibited)
 %Due to fewer modes available for interaction, energy keeps accumulating at certain scales. The moment when energy accumulation is no longer possible, it is cascaded across scales in accordance with the KWT . 
%, such as $k = n\pi/L$ where $n$ is an integer 

Although wave turbulence has been experimentally investigated in various systems~\cite{Falcon2007,LauriePhysRep2012,DeikeJFM2013,HassainiPRE2019,MonsalvePRL2020,FalconARFM2022,RicardPRF2023,LanchonPRF2025}, the influence of finite-system size effects has received little attention, with most experimental studies focusing on hydrodynamic surface wave turbulence~\cite{IssenmannPRE2013,DeikeJFM2015,HassainiPRF2018,CazaubielPRL2019}. The role of the shape of the basin has been addressed~\cite{IssenmannPRE2013} as well as its boundary condition (e.g., absorbing with a beach, or reflecting with a wall)~\cite{DeikeJFM2015}. Finite-system size effects have been explored in a rectangular tank with a movable partition along one direction, showing discrete modes in the confined direction and a continuous spectrum in the perpendicular direction~\cite{HassainiPRF2018}. %The partition was moved to change the aspect ratio, while the length of the tank in the perpendicular direction was kept constant and sufficiently large to avoid discreteness effects in this direction. In this experiment, the spatiotemporal spectrum of gravity-capillary waves along the smaller length exhibits discrete modes, confirming the influence of the confinement, whereas the spectrum in the perpendicular direction was continuous. 
In cylindrical containers, as a consequence of conservation laws, gravity-wave turbulence can sustain three-wave resonant interactions due to spatial confinement (instead of usual four-wave ones)~\cite{MichelPRF2019,DureyJFM2023}. These modified interactions have been evidenced in a gravity-wave turbulence experiment in a high-gravity environment~\cite{CazaubielPRL2019}. %Moreover, the inverse cascade of gravity-wave turbulence was experimentally found to stop well before wave action piles up at the largest scale due to possible finite-basin size effects; the limitation of the inverse cascade resulting from the emergence of dissipative coherent structures (sharp-crested waves)~\cite{FalconPRL2020}. The role of finite-size effects on wave resonant interactions and on wave turbulence thus deserves further studies with accurate space-time measurements.

%Consequently, it is crucial to characterize how finite-size effects influence wave turbulence properties to better understand experimental results in which geometry plays a significant role. 
It is therefore essential to quantify how finite-system size effects shape wave turbulence properties, particularly in experiments where they strongly impact the dynamics. To investigate finite-size effects in gravity-capillary wave turbulence, we use an experimental setup in which the water surface is perturbed locally and erratically by small magnets partially submerged in water. This type of forcing generates waves with statistically homogeneous and isotropic properties while minimizing direct excitation of the container’s sloshing modes, unlike horizontal oscillations of the entire tank~\cite{IssenmannPRE2013,AubourgPRF2016,HassainiPRF2018}, which strongly excite such modes and collinear wave interactions that mask additional confinement effects~\cite{HassainiPRF2018}. Here, we show experimental signatures of finite-system size effects in wave turbulence on the wave spectrum, wave interactions, and typical timescales.

%Hassaini et al. have experimentally demonstrated the finite-size effects on wave turbulence by calculating the 3D spatio-temporal spectra ($S(k_x)$, $S(k_y)$, and $S(\omega)$) of waves in a rectangular tank that had a movable partition along one direction (say $x$). The partition was moved to change the confinement, while the length of the tank in the perpendicular direction ($y$) was kept constant, and it was also sufficiently large to avoid the depletion of modes due to discreteness. The effect of the confinement was reported to be observed in the 3D spectra; the $\omega$-$k_x$ spectra had discrete modes, whereas $\omega$-$k_y$ was continuous.

%Here, we further extend the results of the effects of confinement in wave turbulence. In contrast to Hassaini et al. \cite{HassainiPRF2018}, who employed global forcing on the system, we locally perturb the surface of water with the help of small magnets that are partially submerged in water. This type of forcing generates waves with more homogeneous properties, and to the best of our knowledge, this forcing has not been used in wave turbulence before. }
%Kartashova E, Nazarenko S, Rudenko O. 2008. Resonant interactions of nonlinear water waves in a finite basin. Phys. Rev. E 78:016304
%L’vov VS, Nazarenko S. 2010. Discrete and mesoscopic regimes of finite-size wave turbulence. Phys. Rev. E 82(5):056322
%Pan Y, Yue DKP. 2017. Understanding discrete capillary-wave turbulence using a quasi-resonant kinetic equation. J. Fluid Mech. 816:R1
%While nonlinear effects have been studied extensively,

\section{Theoretical backgrounds}
%\subsection{Linear dispersion relation}
The linear dispersion relation of inviscid linear deep-water waves on the infinite surface of a fluid reads~\cite{Lamb1932}
\begin{equation}
\omega^2 =  gk + \gamma k^3/\rho \ , 
\label{eq1}
\end{equation}
where $\omega \equiv 2\pi f$ is the angular frequency, ${\bf k}$ is the wavevector ($k\equiv2\pi/\lambda= ||{\bf k}||$), $g$ is the acceleration due to gravity, and $\gamma$ and $\rho$ are the surface tension and density of the fluid, respectively. The first term of the right-hand member of Eq.~\eqref{eq1} corresponds to gravity waves, whereas the second term corresponds to capillary waves. The transition between these two pure regimes occurs for $\lambda_{gc}=2\pi \sqrt{\gamma/(\rho g)}$ close to 1~cm for most fluids. Waves with wavelengths near this crossover are gravity-capillary waves which follow the dispersion relationship $\omega(k)$ given by Eq.~\eqref{eq1}. 

%\subsection{Nonlinear wave interactions}
 %Existing waves form new ones of different scales; either several waves add up to make a new wave, or one wave splits into many. 

Weakly nonlinear waves can interact with one another to transfer energy between waves. This nonlinear wave interaction process is the fundamental mechanism of weak-turbulence theory. $N$ nonlinear waves are in {\it resonant} interactions when they simultaneously satisfy the following conditions on angular frequencies $\omega_i$ and  wavevector $\mathbf{k_i}$
\begin{equation}
\omega_1 \pm \omega_2 \pm \cdots \pm \omega_{\it N} = 0  \ \ \mathrm{and} \ \ \mathbf{k_1} \pm \mathbf{k_2} \pm \cdots \pm \mathbf{k_{\it N}} = 0 , \ \ \mathrm{with} \ \ N\geq 3 ,
\label{eq2}
\end{equation}
where each wave $i=1,\ 2, \cdots, N$ follows the linear dispersion relation, $\omega(k)$, and $\omega_i\equiv \omega(||\mathbf{k_i}||)$. $N$ is the minimal number of waves for which Eq.~\eqref{eq2} is satisfied, which thus depends on the geometry and wave dispersion relationship. For instance, when waves propagate in two dimensions (2D), for pure capillary waves or gravity-capillary waves, $N=3$, and, for pure gravity waves,  $N=4$~\cite{Zakharov,NazarenkoBook,GaltierBook}. When the linear dispersion relation is broadened by weakly nonlinear or dissipative corrections, {\it quasiresonant} wave interactions are then possible, so that the two conditions of Eq.~\eqref{eq2} are approximately satisfied, such as
\begin{equation}
\omega_1 \pm \omega_2 \pm \cdots \pm \omega_{\it N} = 0  \ \ \mathrm{and} \ \ \mathbf{k_1} \pm \mathbf{k_2} \pm \cdots \pm \mathbf{k_{\it N}} < \pm \delta \mathbf{k} , \ \ \mathrm{with} \ \ N\geq 3 .
%\omega_1 \pm \omega_2 \pm \omega_3 = 0  \ \ \mathrm{and} \ \ \mathbf{k_1} \pm \mathbf{k_2} \pm \mathbf{k_3}  < \pm \delta \mathbf{k} .
\label{eq3}
\end{equation}
$\delta k=||\delta{\bf k}||$ corresponds to the nonlinear broadening of the dispersion relation. This mismatch can be also written in $\delta \omega$ using Eq.~\eqref{eq1}. Note that the different signs $\pm$ need to be the same in each instance of Eq.~\eqref{eq2} [or Eq.~\eqref{eq3}]. Such resonant and quasiresonant interactions have been experimentally evidenced, in particular, in gravity, gravity-capillary, and capillary wave systems (see review~\cite{FalconARFM2022}). {\it Nonresonant} wave interactions can also occur if the two conditions of Eq.~\eqref{eq2} are fulfilled, but at least one of the involved Fourier modes is not a {\it free wave}, i.e., it does not follow the linear dispersion relation of Eq.~\eqref{eq1}, and is then called a {\it bound wave}. Such bound waves, propagating with the same velocity as a carrier-free wave, lead to several additional branches in the dispersion relation in a $\omega - {\bf k}$ plot~\cite{HerbertPRL2010,CampagnePRF2018,MichelPRF2018}. However, weak-turbulence theory needs resonant interactions (at the lowest nonlinear order) to build wave turbulence~\cite{Zakharov,NazarenkoBook}. In some systems, such as internal gravity waves, {\it nonlocal resonant} interactions (i.e., involving modes with very different wave-vector lengths or frequencies) are essential~\cite{LabarrePRL2025}. Finally, finite-size effects usually manifest through sloshing modes in a bounded container~\cite{IbrahimBook}, giving rise to additional sloshing branches in the $\omega - {\bf k}$ spectrum, as reported for surface waves in a fluid torus~\cite{NovkoskiPRL2021,NovkoskiPRE2023}. Distinct regimes are then expected depending on the ratio of the nonlinear spectral broadening ($\delta k$) to the mode spacing ($\Delta k$): frozen wave turbulence when $\delta k \ll \Delta k$, discrete wave turbulence and mesoscopic wave turbulence when $\delta k \sim \Delta k$, and continuous wave turbulence when $\delta k \gg \Delta k$.

\begin{figure}[t!]
    \centering
       \includegraphics[width=0.71\linewidth]{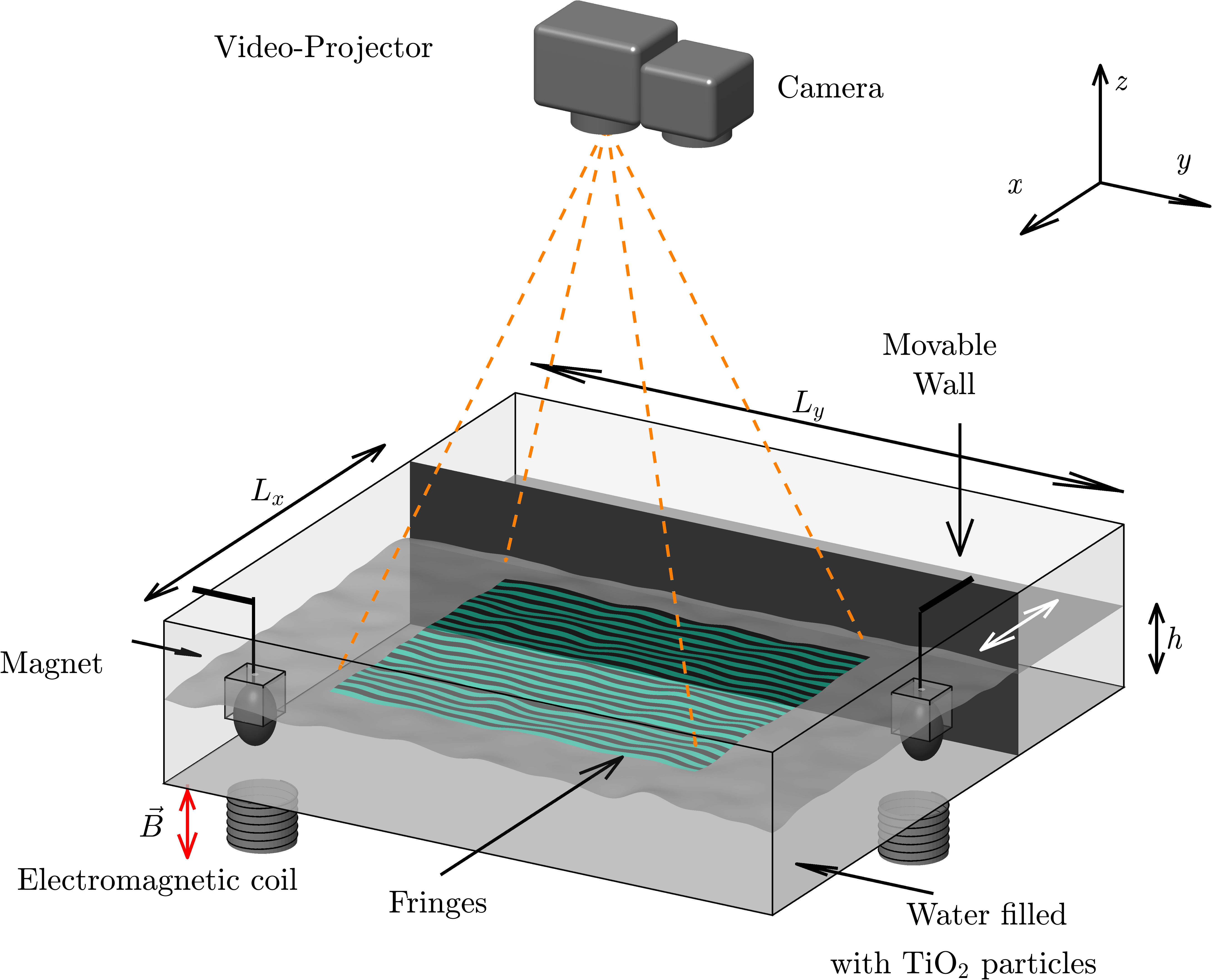}
       \put(-300,190){\fontsize{12}{14}\selectfont (a)}
       \hspace{1.5cm}
 \raisebox{3.7cm}{{\includegraphics[width=0.1\textwidth]{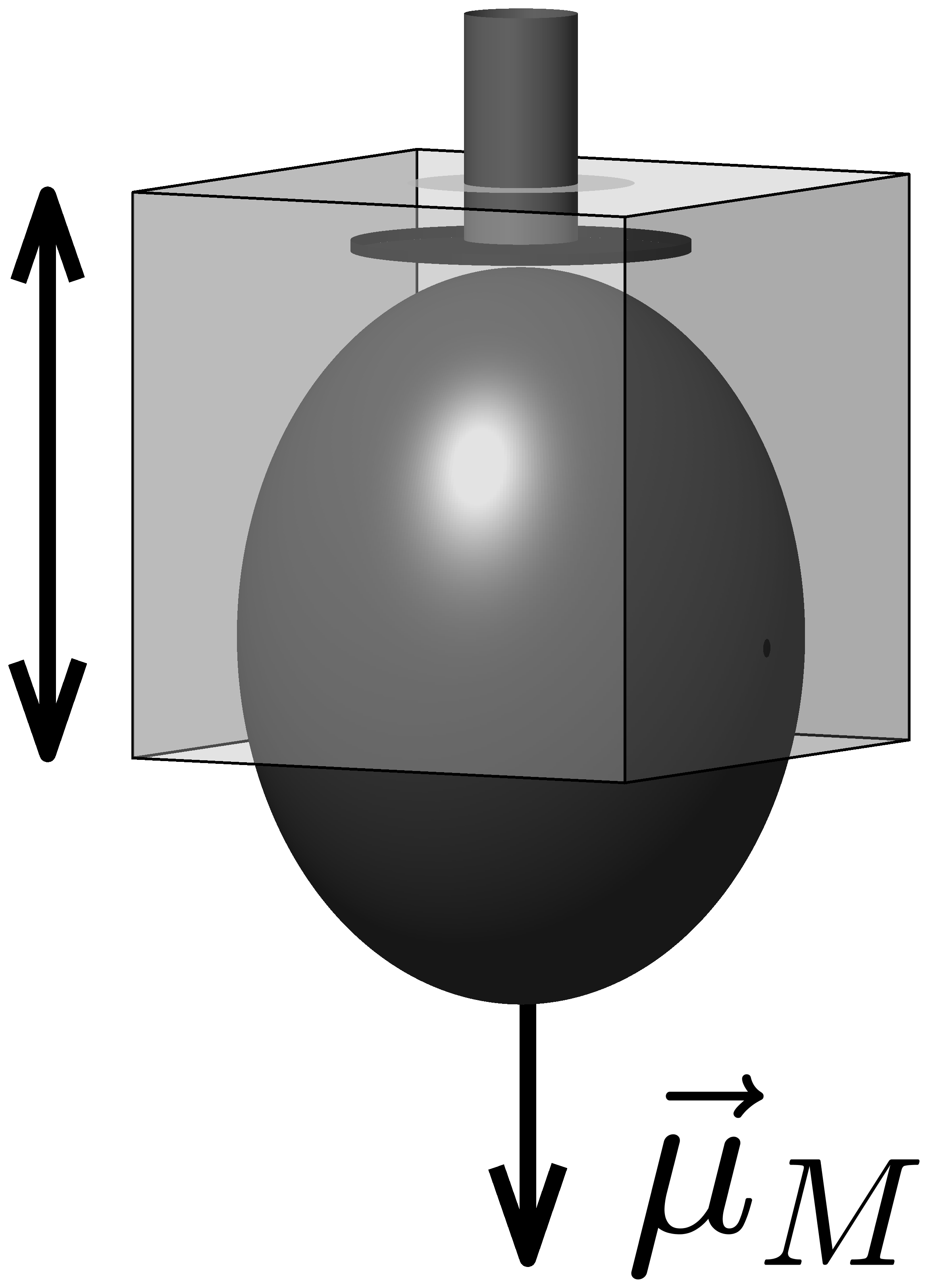}}}
\put(-75,150){\fontsize{10}{12}\selectfont 1cm }
 \put(-30,195){\fontsize{12}{14}\selectfont (b)}
\put(-50,180){\fontsize{10}{12}\selectfont Magnet-case}
\caption{(a) Schematic diagram of the experimental setup used to generate waves in a confined environment.  Random waves are generated by two magnets partially submerged in the fluid, and energized by electromagnetic coils located beneath the container. A movable partition changes the container aspect ratio by reducing the length in the $x$ direction. A fringe pattern is projected on the surface of water, and the spatiotemporal deformations of the fringes are recorded using a high-speed camera. (b) Schematic diagram of the case used to suspend a magnet.}
\label{setup}
\end{figure}

% \raisebox{3.7cm}{{\includegraphics[width=0.1\textwidth]{Fig1a.png}}}
%\put(-75,150){\fontsize{10}{12}\selectfont 1cm }
%\put(-30,195){\fontsize{12}{14}\selectfont (a)}
%\put(-50,180){\fontsize{10}{12}\selectfont Magnet-case}
%\hspace{1cm}
%    \includegraphics[width=0.7\linewidth]{Fig1b.png}
%    \put(-205,190){\fontsize{12}{14}\selectfont (b)}

\section{Experimental setup}\label{expsetup}
A schematic diagram of the experimental setup used to investigate finite-size effects in gravity-capillary wave turbulence is presented in Fig.~\ref{setup}a. The experiments are conducted in a rectangular transparent Plexiglass container, with the length along the confined direction ($x$) adjustable in the range $L_x \in [5, 100]$~cm using a movable wall, while the unconfined direction is fixed at $L_y=50$~cm.  The aspect ratio $\mathrm{AR} \equiv L_y/L_x $ is thus changed by one and a half decades in the range $\mathrm{AR} \in [0.5,10]$. The distilled-water depth is maintained at $h=2$~cm to ensure a deep-water regime ($\lambda < 2\pi h$). Surface waves are generated by two prolate elliptical rare-earth magnets encased in a PTFE coating (1~cm in diameter, 1.5~cm in length, magnetic moment $\mu_m=0.8$~A~m$^{2}$~\cite{Fisherbrand}) and actuated by an electromagnetic coil~\cite{coils} positioned beneath the container, directly below each magnet. Each magnet is suspended by a flexible, nonextensible 1-mm diameter cotton cord at the two diagonally opposite corners of the container. The string lengths are adjusted so that the magnets remain submerged just below the water surface. The erratic motion of each magnet is induced by an AC vertical magnetic field $\vec{B}(t)$ generated by each electromagnetic coil, driven with a random noise current in frequency ($2\pm 0.5$~Hz) and amplitude (see movie {\em MagneticRandomForcing.mp4} in Supp. Mat.~\cite{SuppMat}). As the response of each suspended magnet to such a magnetic field is not the same, this forcing generates waves with uncorrelated properties. Typically, the time-dependent torque $\vec{\Gamma}(t)=\vec{\mu}_{M}\times \vec{B}(t)$ imposed by the coil on the magnet generates its erratic motion~\cite{FalconEPL13,FalconPRF2017,GorcePRE2023}. This wave-forcing method is novel and was previously developed by our group to drive, randomly in space and time, a granular gas
~\cite{FalconEPL13,FalconPRF2017,GorcePRE2023} or three-dimension (3D) hydrodynamics turbulence~\cite{CazaubielPRF2021,GorcePRL2022,GorcePRL2024}. The forcing amplitude of the magnets is controlled by the current strength, $I\in[3,7]$~A, flowing into the coils from a 2~kW power supply (QualitySource PA2000AB). Each magnet is housed inside a 3D-printed cubic shell, 1~cm in size (see Fig.~\ref{setup}b). The top face of the cube features a circular hole that allows a cylindrical shell to rotate freely. A flexible, nonextensible string is attached to this cylindrical shell to suspend the magnet assembly. The free rotation of the cylinder within the hole of the case ensures that there is no torsion in the string when the coils energize the magnets. A crucial feature of this forcing is that it does not directly excite the container's sloshing modes, unlike when the entire container is horizontally oscillated~\cite{IssenmannPRE2013,AubourgPRF2016,HassainiPRF2018}, thus potentially masking additional confinement effects when investigated~\cite{HassainiPRF2018}. Our electromagnetic forcing also leads to a homogeneous wave field (see below) and mainly forces the free surface rather than the bulk of the fluid (as is the case when flaps are used). %The forcing amplitude is controlled by the random noise amplitude. 

The fully space-and-time resolved surface wave-height field, $\eta(x,y,t)$, is measured using Fourier transform profilometry (FTP)~\cite{Takeda,Cobelli}. A sinusoidally-coded fringe pattern is projected by a full-HD video projector (Epson EH-TW3200) on the water surface. When subjected to forcing, the deformations of the fringes are recorded using a high-speed camera (Phantom v10), positioned above the fluid, at 120 fps and a spatial resolution of $112 \times 1800$ pixels$^2$ corresponding to a surface of $\mathcal{S}=1.7 \times 27.4$~cm$^2$ (along the $x$ and $y$ directions) around the center of the container. For better comparison, we kept the spatial resolutions in $x$- and $y$-directions constant, regardless of the confinement size. %rather than a confinement-dependent spatial resolution in the $x$-direction.
It is ensured that the magnets were not in the field of view of the camera. To enhance water optical diffusivity and clearly project the fringe pattern onto the water surface, a controlled amount of micrometric titanium dioxide (TiO$_2$) particles is added to water~\cite{TiO2}.  Previous studies have confirmed that TiO$_2$ particles do not alter the water surface tension and viscosity~\cite{Przadka2012}, ensuring that the measurements remain unaffected by the seeding particles. Due to the deformation of the surface in the presence of waves, the phase information of the projected fringes changes with respect to the fluid surface at rest (i.e., without forcing).
%In each set of experiments, the water surface at rest (i.e., without forcing) is first recorded (reference image) to get the phase information of the projected fringes. Subsequently, the magnets force the free surface, and the surface deformations are again recorded.  Due to the deformation of the surface in the presence of waves, the phase information of the light reaching the camera changes with respect to the reference image. The Fourier transform of the reference image is then subtracted from that of the surface with waves at any instant; the corresponding phase difference information can be extracted. 
This map of phase differences is then used to calculate the wave-height field, $\eta(x,y)$, at each time step. Then, a (2+1)D Fourier transform provides the wave-height field in the Fourier space, $\hat{\eta}(k_x,k_y,\omega)$, thus leading to the 3D spectrum of the wave height $S_{\eta}(k_x,k_y,\omega)\equiv |\hat{\eta}(k_x,k_y,\omega)|^2/(\mathcal{ST})$ with $\mathcal{T}=60$~s the acquisition time. The spatial periodicity of the projected fringe pattern is 2~mm. Note that, for convenience, we compute the spectrum of the vertical-velocity field of the waves, $v(x,y,t) = \partial \eta(x,y,t)/\partial t$, noted  $S_{v}(k_x,k_y,\omega)$. For all experiments, the wave steepness is weak enough to ensure that waves remain weakly nonlinear. The wave steepness is indeed varied in a range $\epsilon \in [0.3, 2]$\%, which is experimentally quantified by the wave mean slope as $\epsilon \equiv \overline{(\int_\mathcal{S} ||\nabla \eta(x,y,t)||^2 \,dx\,dy/\mathcal{S})^{1/2}}$ where $\overline{\ \cdot \ }$ stands for a temporal average. The typical rms wave height is of the order of 0.2~mm. %The FTP technique cannot resolve waves with wavelengths shorter than this fringe spacing. 

\begin{figure}[b!]
\includegraphics[height=9cm]{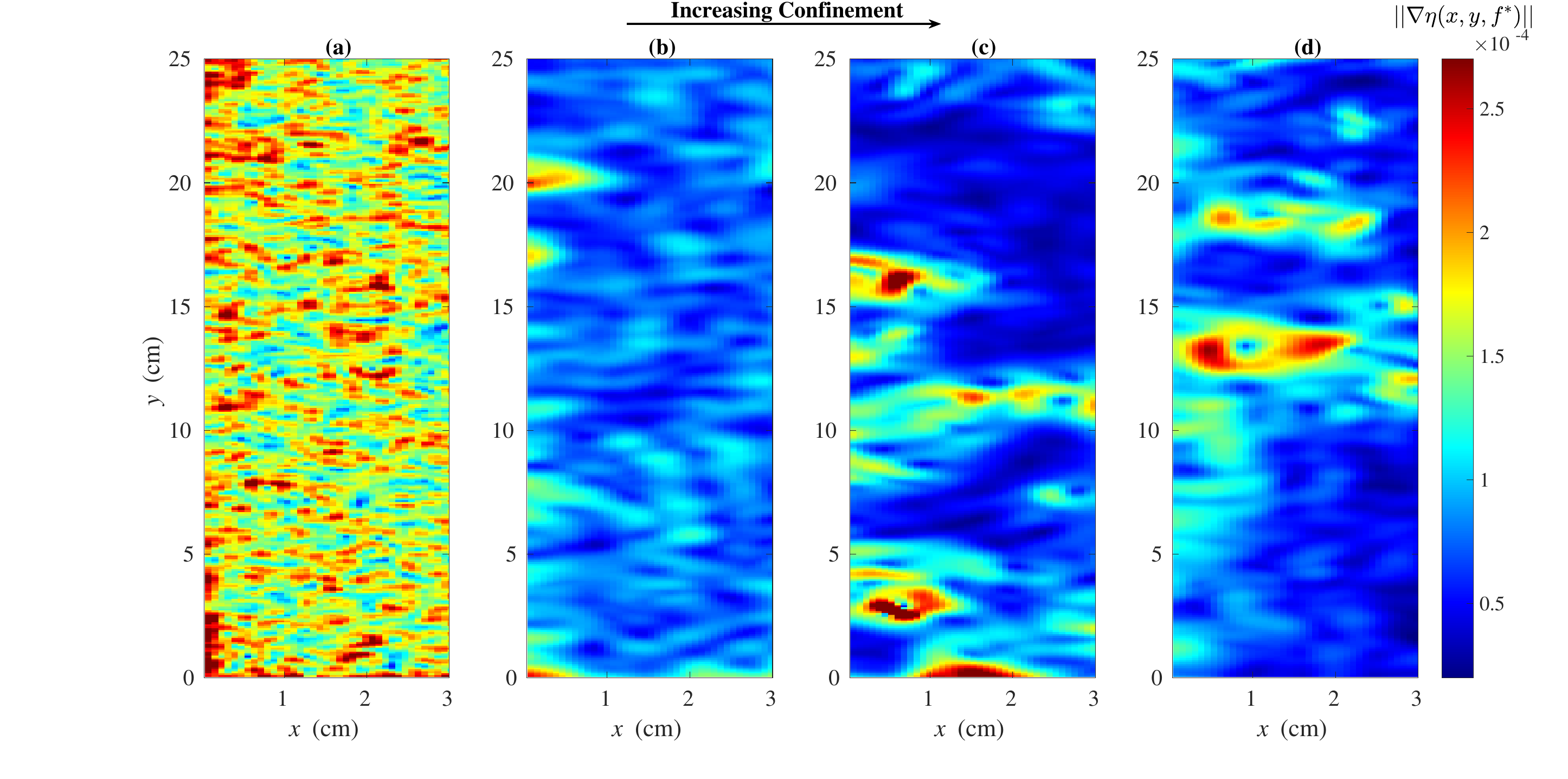}%gradient.eps} %width=9cm,
\caption{Spatial profile of the wave field gradient $||\boldsymbol{\nabla} \eta(x,y,f^*)||$, near the center of the container, for different confinements $L_x$ (at fixed $L_y=50$~cm) corresponding to different container aspect ratios $\mathrm{AR}\equiv L_y/L_x$: (a) $L_x=100$~cm (unconfined case $\mathrm{AR}=0.5$), (b) $L_x=11$~cm ($\mathrm{AR}=4.54$), (c) $L_x=8$~cm ($\mathrm{AR}=6.25$), and (d) $L_x=5$~cm ($\mathrm{AR}=10$). Only one single Fourier mode ($f^*\equiv \omega^*/2\pi=10$~Hz) is selected to estimate the wave field gradient. Random forcing: $2\pm 0.5$~Hz. $\epsilon=2$\%.} \label{fig1}
\end{figure}

\section{Finite-size effects}
\subsection{Gradient of the wave field}
To first identify the finite-size effects of the confinement, we display in Fig.~\ref{fig1} the experimental maps of the local gradients of the wave field for different confinements $L_x$ (at fixed $L_y=50$~cm) corresponding to different container aspect ratios AR. For simplicity, we select only one single Fourier mode ($f^*=10$~Hz) to show the wave field gradient $||\boldsymbol{\nabla} \eta(x,y,f^*)||$, corresponding to the local slope at each point ($x$, $y$) as $\boldsymbol{\nabla} \eta(x,y) \equiv \frac{\partial \eta}{\partial x}\vec{i} + \frac{\partial \eta}{\partial y}\vec{j}$. The value of $f^*$ is chosen within the inertial range of the wave-turbulence cascade (see Sects.~\ref{STspectra} and~\ref{Tspectra}). For the unconfined case [Fig.~\ref{fig1}(a)], the wave field is roughly homogeneous, mainly involving centimetric wavelengths as expected by $\lambda(f^*)$ using Eq.~\eqref{eq1}. %of the order of $\lambda(f^*)\simeq2$~cm using
On the contrary, with confinement [Fig.~\ref{fig1}(b-d)], larger-scale waves are more present in the system than in the unconfined case [Fig.~\ref{fig1}(a)].  This phenomenon arises from the emergence of multiple branches in the dispersion relation caused by the finite size of the container. We will present in the next Sect.~\ref{STspectra}, spatiotemporal spectra at different confinements showing that larger-scale waves are generated (at a fixed frequency) than those coming from the unconfined dispersion relation of Eq.~\eqref{eq1}. This will also correspond to a transition from a continuous wave turbulence cascade (without confinement) to a discrete wave turbulence (strong confinement) (see Sect.~\ref{Tspectra}). 

\begin{figure}[b!] 
%\begin{centering}
\includegraphics[width=8cm]{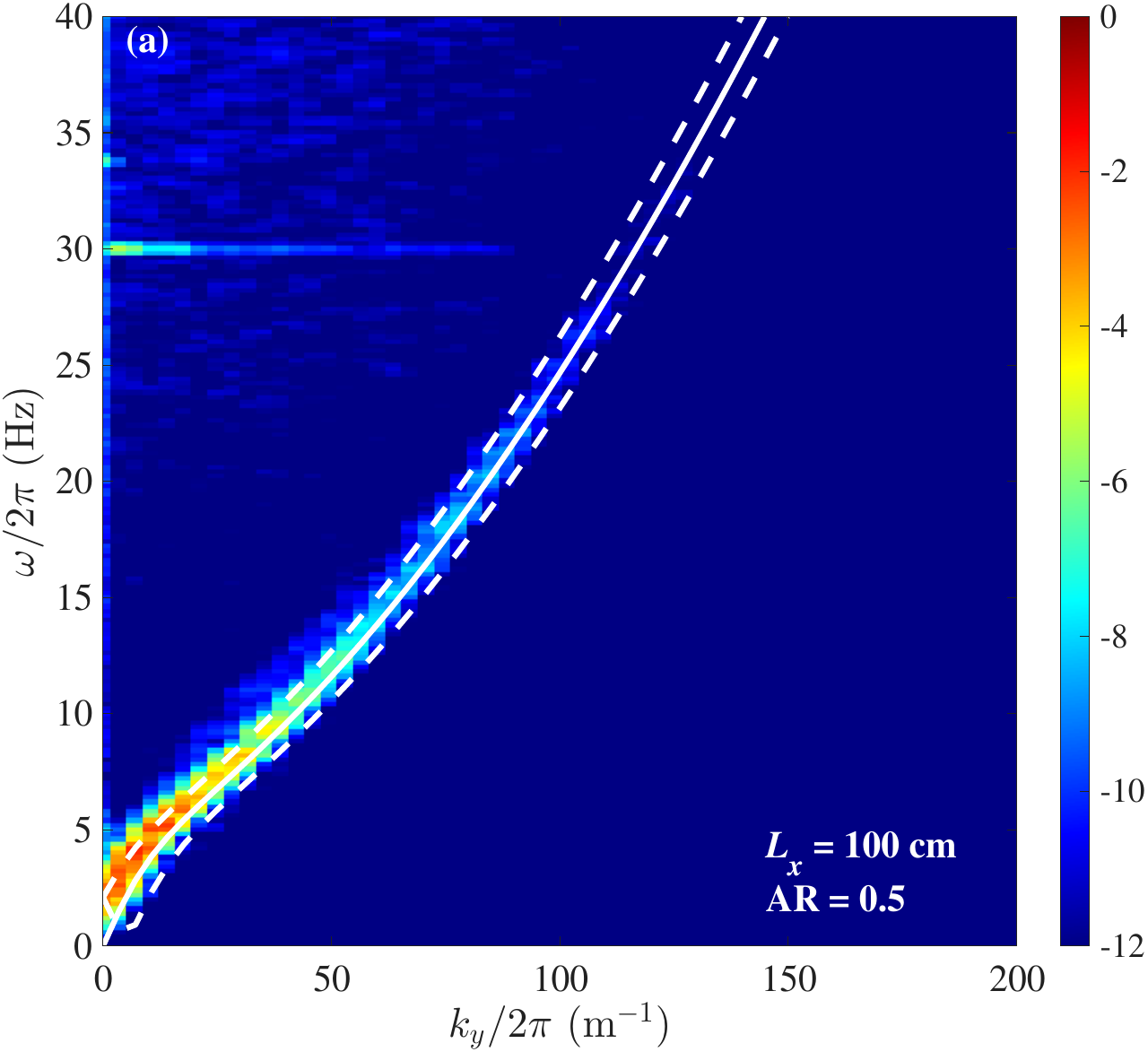} %2a.eps ,height=7cm
\includegraphics[width=8cm]{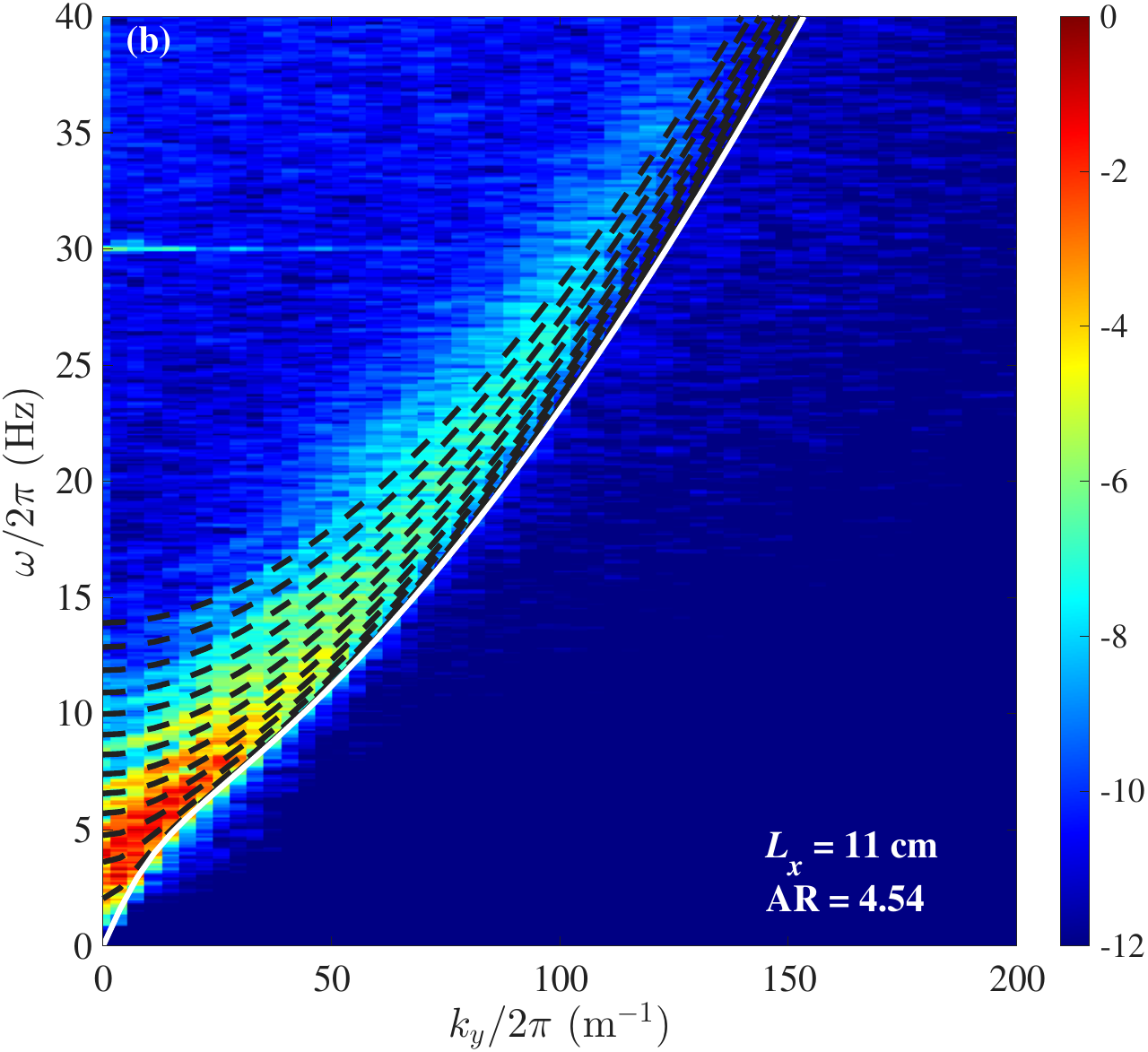}
\includegraphics[width=8cm]{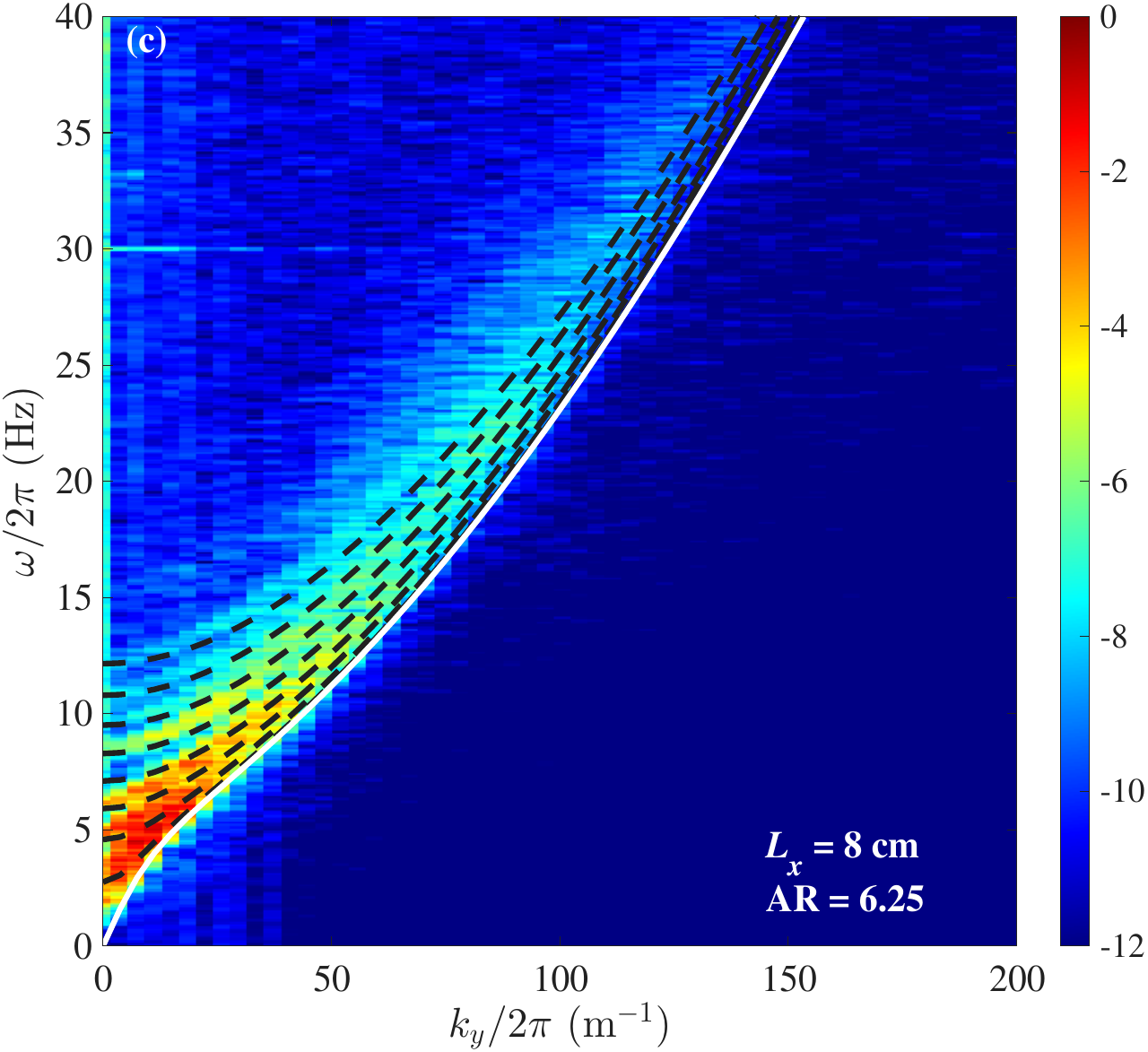}
\includegraphics[width=8cm]{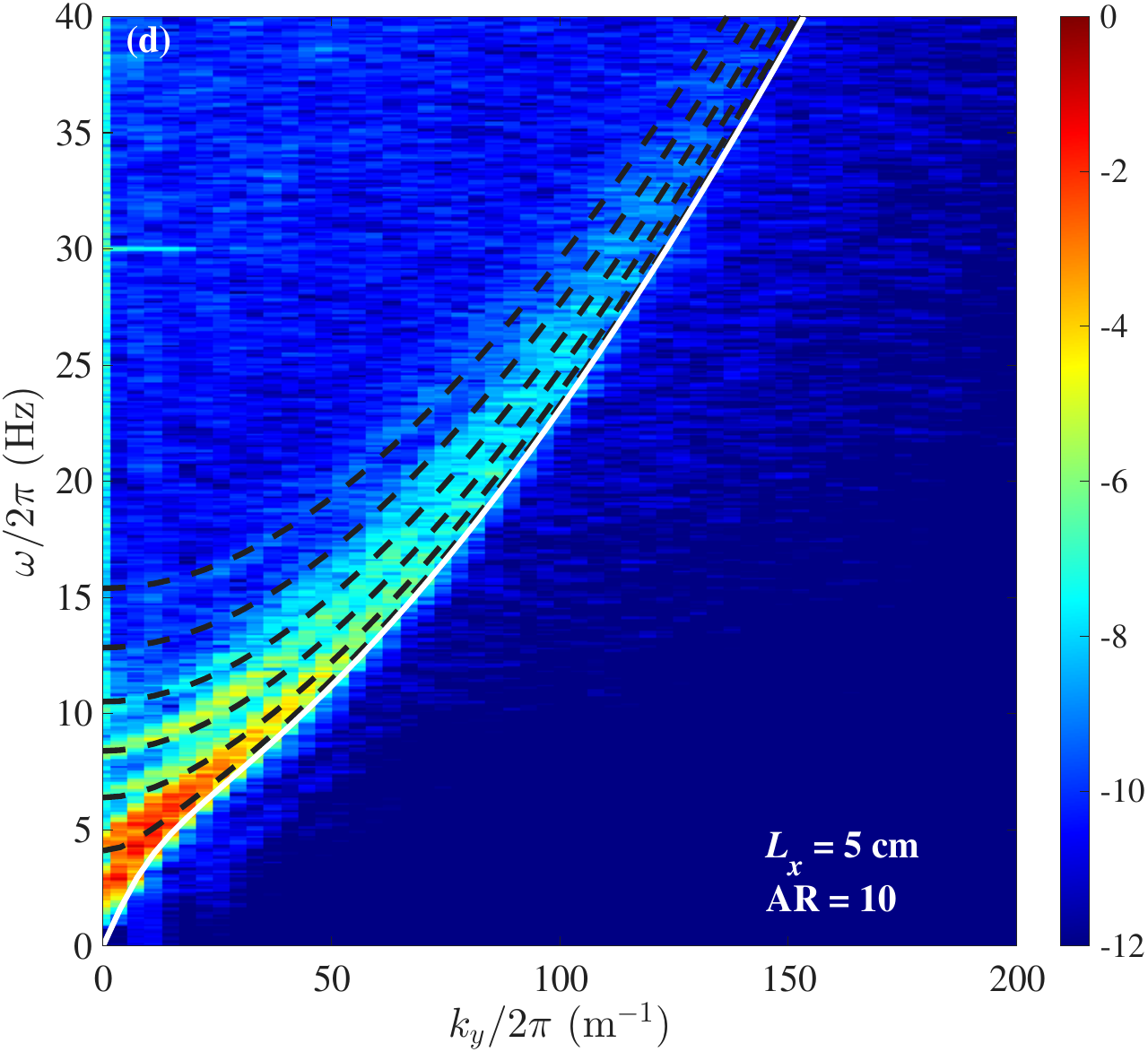}
\caption{Spatiotemporal spectra of the wave vertical-velocity field, $S_v(k_y,\omega)$, for different confinements $L_x$ at fixed $L_y=50$~cm corresponding to different container aspect ratios $\mathrm{AR}\equiv L_y/L_x$: (a) $L_x=100$~cm (unconfined case $\mathrm{AR}=0.5$), (b) $L_x=11$~cm ($\mathrm{AR}=4.54$), (c) $L_x=8$~cm ($\mathrm{AR}=6.25$), and (d) $L_x=5$~cm ($\mathrm{AR}=10$). Log-scale colorbar. Random forcing: $2\pm 0.5$~Hz. $\epsilon=2$\%. The white curve represents the theoretical dispersion relation of linear waves of Eq.~\eqref{eq1}, and the black dashed lines represent different sloshing modes of Eq.~\eqref{eq6}. Dashed white lines in (a) correspond to the nonlinear broadening of the dispersion relation $\omega(k\pm \delta k)$ with $\delta k/(2\pi)=5$~m$^{-1}$.} 
\label{fig2}
\end{figure}

% corresponding to different container aspect ratios AR
\subsection{Spatiotemporal spectrum of the wave field}\label{STspectra}
Figure~\ref{fig2}(a-d) shows the spatiotemporal spectra of the wave vertical-velocity field, $S_v(k_y,\omega)$, along the $y$-direction (at $k_x=0$), for different confinements $L_x$ (at fixed $L_y=50$~cm). Figure~\ref{fig2}(a) corresponds to the spectrum for the unconfined case, i.e., for the largest container ($\mathrm{AR}=0.5$), in which almost no finite-size effects occur. Indeed, the wave energy is observed to spread from the forcing scales ($\lesssim 2$~Hz) down to smaller scales, over more than one decade in frequency, around the theoretical dispersion relation of Eq.~\eqref{eq1} (see white curve), thus indicating a wave turbulence cascade~\cite{FalconARFM2022}. Homogeneity is preserved for this unconfined case, as a similar spectrum is found in the $x$-direction. When the aspect ratio is strongly increased [Fig.~\ref{fig2}(b-d)], this wave turbulence regime is preserved in the unconfined direction (and still well described by the white curve), but finite-size effects now occur. Indeed, a significant amount of energy is now present outside the dispersion relation, in the form of several branches as shown in Fig.~\ref{fig2}(b-d). Due to the finite system size, only waves whose wavelengths are integer fractions of the system size can exist in the confined direction, i.e., discrete modes $k_x=n\pi/L_x$, where $n$ is an integer, and $L_x$ is the length of the container in the confined direction. The dispersion relation of Eq.~\eqref{eq1} is thus affected by finite-size effects as
\begin{equation}
\omega^2 =  gk_n + \gamma k_n^3/\rho \ \ \ \mathrm{with} \ \ \ k_n = \sqrt{\left(\frac{n\pi}{L_x}\right)^2 + k_y^2}\ .
\label{eq6}
\end{equation}
The black-dashed lines in Fig.~\ref{fig2}(b-d) correspond to Eq.~\eqref{eq6} for $n>0$ and different values of $L_x$. Each branch corresponds to one of the modes given by the value of $n=1$, 2, 3, \ldots.  These branches appear in the spectral energy along the unconfined direction, when the discrete modes ($k_x=n\pi/L_x$) in the confined direction contribute enough to modify the wavenumber modulus in Eq.~\eqref{eq6}. These branches are thus sloshing branches. When the confinement of the container is relaxed ($L_x$ is increased), we observe two additional effects: (i) the number of visible branches increases [e.g., from six in Fig.~\ref{fig2}(d) to thirteen in Fig.~\ref{fig2}(b)], and (ii) the sloshing branches are getting closer together and become denser as they approach the dispersion relation. This is a consequence of the $k_x=n\pi/L_x$ term in Eq.~\eqref{eq6} which governs the spacing between the discrete modes. Moreover, for fixed $k_y$ and $n$, the frequency of the discrete mode decreases as $L_x$ increases and approaches the forcing frequency ($\lesssim 2$~Hz). When $L_x$ is large enough that there is almost no confinement [Fig.~\ref{fig2}(a)], sloshing branches thus vanish, and only the wave turbulence regime remains. This analysis was focused on the $y$-direction, where the spatial resolution enables highly accurate spectral measurements, particularly valuable under strong confinement conditions. %Note that this analysis was focused on the $y$-direction, as the resolution along the confined $x$-direction does not yield sufficiently accurate spatial spectra under strong confinement.

%Note that the discrete spectrum along the confined $x$-direction is not presented here, as the resolution of the images in this direction was not sufficiently high to get good enough spatial spectra for the strongest confinements.
 \begin{figure}[ht!]
\includegraphics[height=7.5cm]{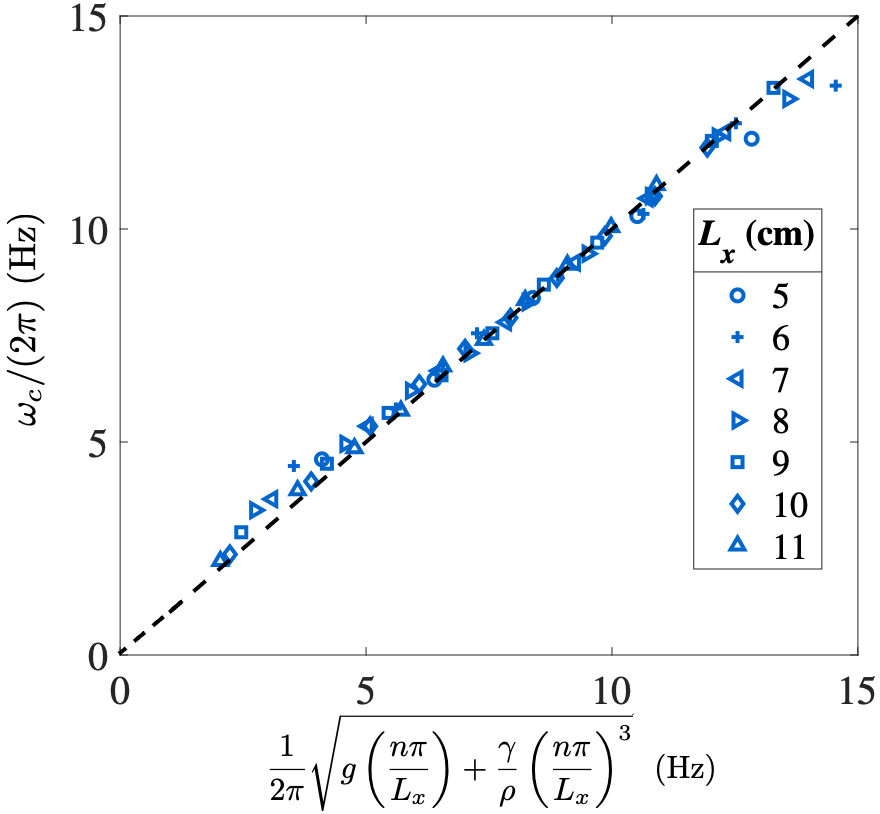}%3.eps} %width=8.5cm,
\caption{Experimental low-frequency cutoffs of sloshing modes [from Fig.~\ref{fig2}(b-d)], for different confinements (see symbols) and all tested wave steepnesses $\epsilon$, as a function of the theoretical prediction, $f_{\mathrm{theo}}$, of Eq.~\eqref{eqfcutoff} with $n=1$, 2, 3, $\ldots$. The black dashed line has a slope of one.} %$f_{\mathrm{theo}}= \sqrt{g\left(\frac{n\pi}{L_x}\right) + \frac{\gamma}{\rho} \left(\frac{n\pi}{L_x}\right)^3}/(2\pi)$ 
\label{fig3}
\end{figure}
\subsection{Sloshing modes}
The intersection of a sloshing branch with the $\omega$-axis in Fig.~\ref{fig2}(b-d) provides its experimental low-frequency cutoff, $\omega_c/(2\pi)$. For all branches and all confinements, these cutoff frequencies are plotted in Fig.~\ref{fig3}, as a function of their theoretical value, 
\begin{equation}
f_{\mathrm{theo}}= \frac{1}{2\pi}\sqrt{g\left(\frac{n\pi}{L_x}\right) + \frac{\gamma}{\rho} \left(\frac{n\pi}{L_x}\right)^3},
\label{eqfcutoff}
\end{equation}
obtained by substituting $k_y=0$ in Eq.~\eqref{eq6} for each value of $n$. We observe that the theoretical predictions are in perfect agreement with the experimental values, as shown by the dashed-line slope of one, and independent of the tested wave steepnesses and confinements.
%that the variation of the rescaling is independent of the system's parameters after they are rescaled, indicating a universal behavior of the sloshing branches across different confinements.

\begin{figure}[b!]
\includegraphics[width=8.5cm]{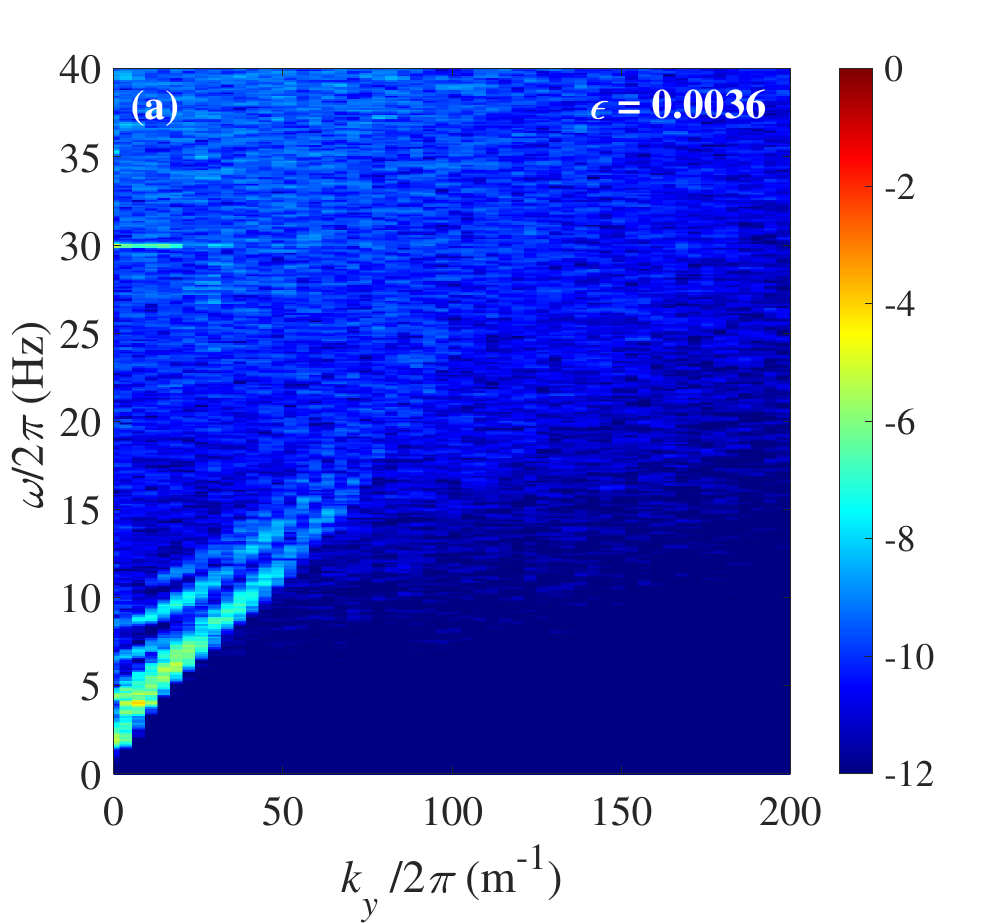} %height=8cm
\includegraphics[width=8.5cm]{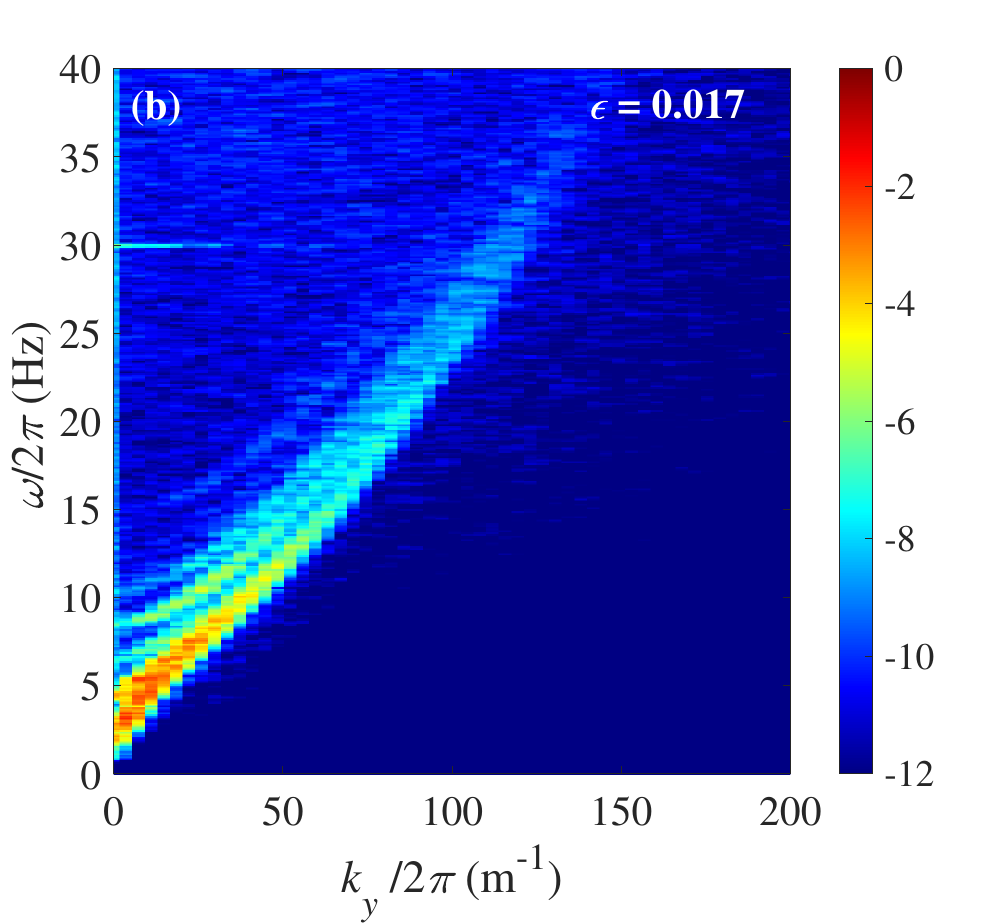}
 \includegraphics[width=8.5cm]{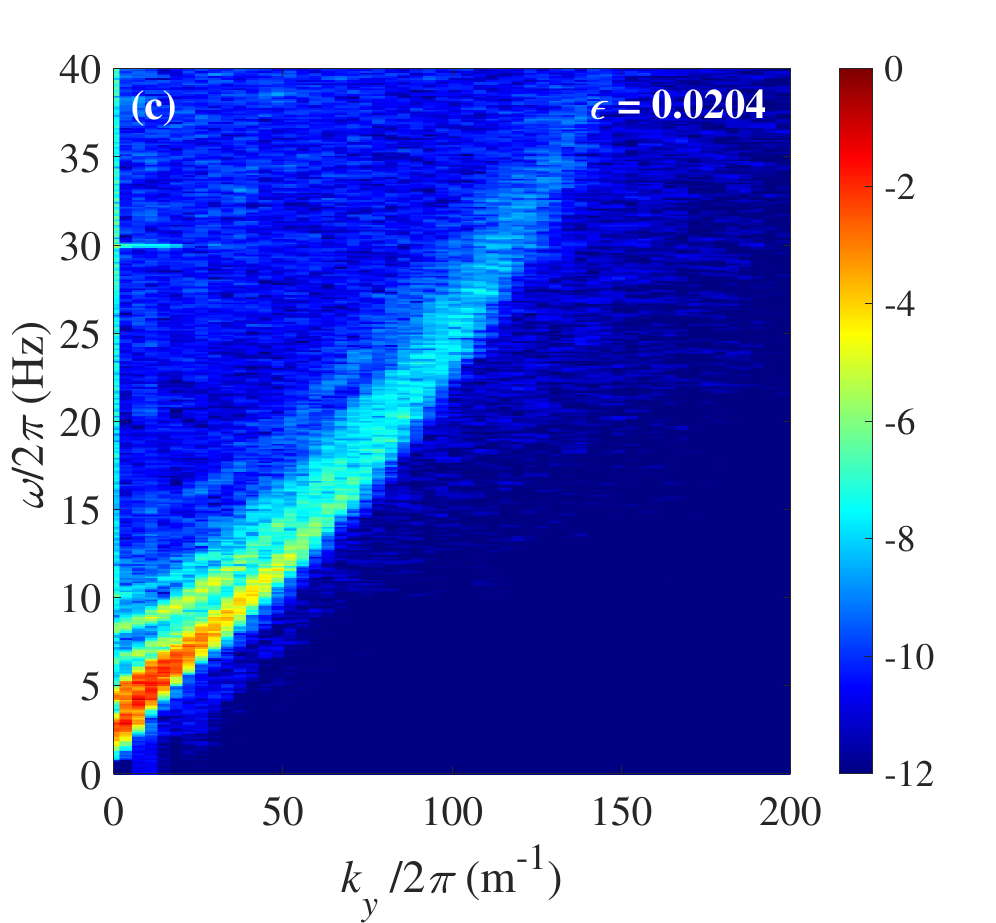}
\caption{Spatiotemporal spectra of the wave vertical-velocity field, $S_v(k_y,\omega)$, for the strongest confinement ($L_x=5$~cm) and different wave steepnesses: (a) $\epsilon=0.36$\%, (b) $\epsilon=1.7$\%, and (c) $\epsilon=2$\% [same as Fig.~\ref{fig2}(d)]. Log-scale colorbar. Random forcing: $2\pm 0.5$~Hz. Experimental data in Fig. 5(c) are the same as in Fig. 3(d).}
\label{fig4}
\end{figure}

\subsection{Role of the wave steepness}
Let us now focus on the effect of the wave steepness $\epsilon$ at a fixed confinement. The spatiotemporal spectra of the wave vertical-velocity field, $S_v(k_y,\omega)$, are shown in Fig.~\ref{fig4} for different wave steepnesses $\epsilon$, and for the strongest confinement ($L_x=5$~cm). As explained in Sect.~\ref{expsetup}, different wave steepnesses are achieved by forcing magnets using different strengths of the random current feeding the coils. We keep the same logarithmic scale for the wave-energy magnitude in Fig. \ref{fig4} to focus on the emergence of sloshing modes and the wave turbulence cascade. At the lowest $\epsilon$, part of the energy injected at large scales spreads towards small scales around the dispersion relation, while a few branches of sloshing modes start to emerge. At increased nonlinearity (but still weak), the main branch broadens, and the wave-turbulence energy cascade is visible up to much smaller scales, while additional sloshing branches appear. These observations highlight the onset of finite-container size effects for weak increasing nonlinearities. %%The finite-size effect of the container thus occurs even for very weak nonlinearities.  %, and related it to the growing importance of the nonlinear contribution in the wave interaction dynamics.

\begin{figure}[t!]
\includegraphics[width=8.5cm]{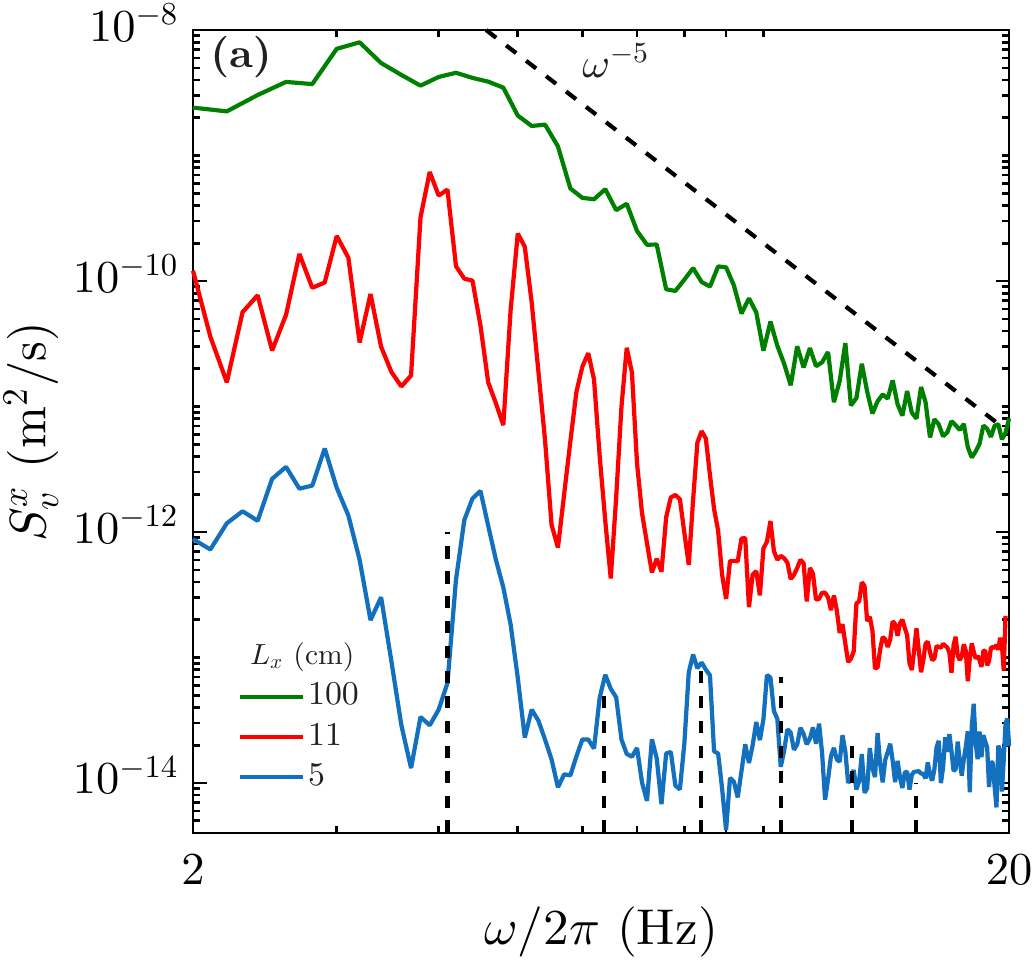}%5a.eps}[width=8.5cm,height=7cm]
\hspace{0.5cm}
\includegraphics[width=8.5cm]{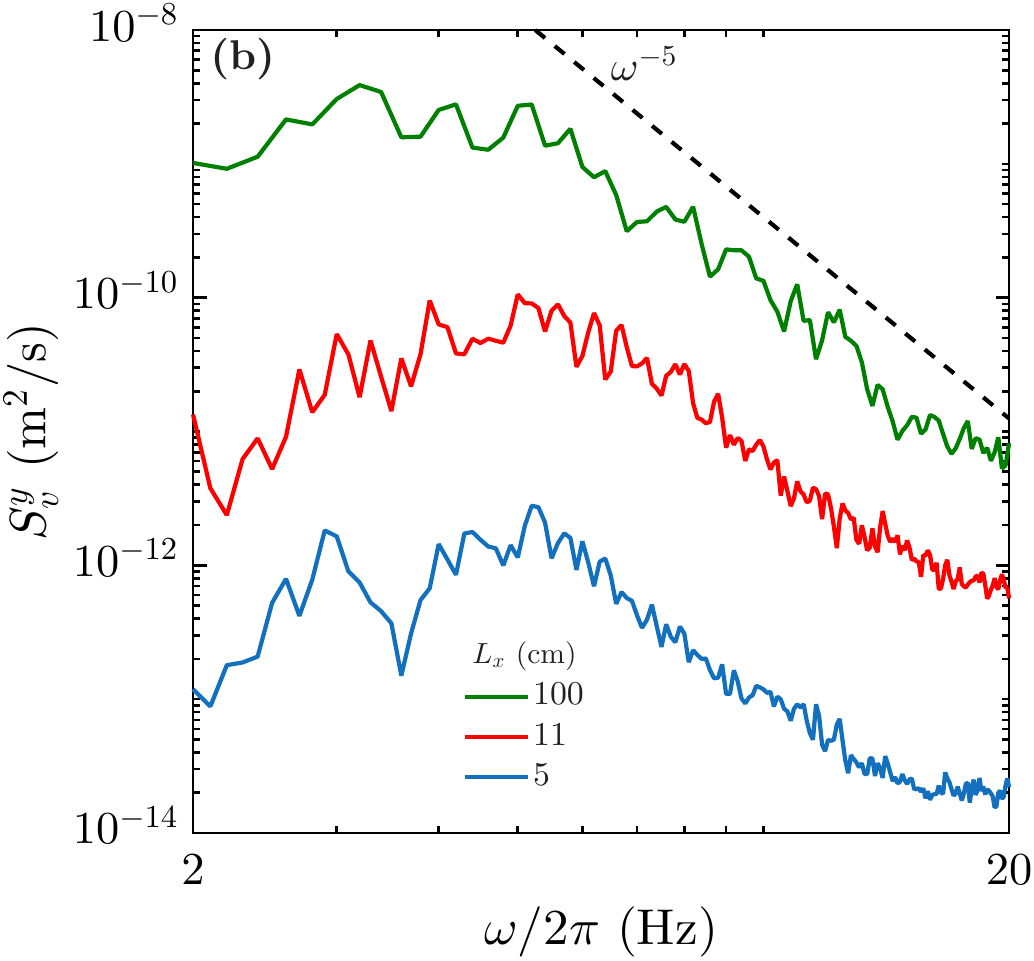}%5b.eps}[width=8.5cm,height=7cm]{
\caption{Frequency spectra of the wave vertical velocity along (a) the confined $x$-direction, $S_v^x(\omega)$, and (b) the unconfined $y$-direction, $S_v^y(\omega)$, for different confinements $L_x\in[5,11]$~cm and the unconfined case $L_x=100$~cm (from bottom to top). Spectra are vertically shifted for clarity. Random forcing: $2\pm 0.5$~Hz. For the strongest confinement ($L_x=5$~cm), vertical dashed lines in (a) represent the theoretical cutoff frequencies of sloshing modes of Eq.~\eqref{eq6} with $k_y=0$, with $n=1$ to 6. }
\label{fig5}
\end{figure}

%\begin{figure}[ht!]
%\includegraphics[width=8.5cm,height=7cm]{6.eps}
%\caption{Frequency spectrum of the wave vertical velocity along (a) the confined $x$-direction for the strongest confinement ($L_x=5$~cm). Gray dashed vertical lines represent the cut-off frequencies of the corresponding sloshing modes from Eq.~\eqref{eq6} with $k_y=0$ with $n=1$ to 6.}
%\label{fig6}
%\end{figure}

\subsection{Frequency spectrum}\label{Tspectra}
We now compute the frequency spectra of the wave vertical velocity, $S_v(\omega)$, for different confinements. The frequency spectra along the confined direction, $S_v^x(\omega)$, and along the unconfined direction, $S_v^y(\omega)$, are plotted in Fig.~\ref{fig5} to compare the container finite-size effects on wave directions. The frequency spectrum is obtained from the spatiotemporal spectrum $S_v(k_x,k_y,\omega)$ as $S_v^y(\omega) \equiv \int S_v(k_x=0,k_y,\omega)dk_y$ for the unconfined direction, and $S_v^x(\omega) \equiv \int S_v(k_x,k_y=0,\omega)dk_x$ for the confined direction. The Fourier spectrum in the confined direction, $S_v^x(\omega)$ in Fig.~\ref{fig5}(a), is completely different from that in the unconfined direction, $S_v^y(\omega)$, in Fig.~\ref{fig5}(b). Remarkably, $S_v^x(\omega)$ in Fig.~\ref{fig5}(a) shows a transition from a continuous spectrum for the unconfined case ($L_x=100$~cm) to a discrete spectrum for confined cases ($L_x=5$  and 11~cm), whereas the spectrum in the unconfined direction, $S_v^y(\omega)$, is continuous regardless of the confinement level. 

The distinct peaks present in the $S_v^x$ spectrum, in Fig.~\ref{fig5}(a), well correspond to the theoretical cutoff frequencies of the sloshing branches of Eq.~\eqref{eq6} with $k_y=0$ (see vertical dashed lines) due to the confinement. These peaks become less separated and shift to lower frequencies when the confinement is relaxed, as expected by Eq.~\eqref{eq6}. These observations are consistent with the one reported experimentally for a horizontal forcing~\cite{HassainiPRF2018} and with predictions~\cite{KartashovaEPL2009,NazarenkoJSM2006}. %shows also vertical dashed lines corresponding to the theoretical cut-off frequencies of the sloshing branches of Eq.~\eqref{eq6} with $k_y=0$. We observe a rather good agreement between these theoretical values and the peaks of the experimental spectrum $S_v^x(\omega)$ for the strongest confinement ($L_x=5$~cm).

%Moreover, we observe, in $S_v^x(\omega)$ in Fig.~\ref{fig5}(a), a transition from a continuous spectrum (for the unconfined case $L_x=100$~cm) to a discrete spectrum for confined cases ($L_x=5$  and 11).  
%S_v^x(\omega)$ is replotted in for the strongest confinement ($L_x=5$~cm) along with vertical dashed lines corresponding to the cut-off frequencies of the sloshing branches of Eq.~\eqref{eq6} with $k_y=0$. We observe a rather good agreement between the theoretical values and experimental results.%explain the slight departure for $n=1$: too close to the forcing harmonics frequency

On the contrary, the $S_v^y(\omega)$ spectrum, in Fig.~\ref{fig5}(b), is continuous and displays a frequency power-law cascade corresponding to a gravity-capillary wave turbulence cascade.  The best power-law fit typically gives $S_v^y(\omega) \sim \omega^{-5}$ for the wave vertical-velocity spectrum, meaning thus to a steeper spectrum for wave heights $\eta$, $S_{\eta}^y(\omega) \sim \omega^{-7}$. This exponent thus deviates significantly from weak-turbulence predictions in the pure gravity regime, $S^g_{\eta}\sim \omega^{-4}$~\cite{ZakharovGrav1966}, and in the pure capillary regime, $S^c_{\eta}\sim \omega^{-17/6}$~\cite{ZakharovCapi1967}. The deviation of the power-law exponent of the frequency spectrum in the gravity regime has previously been reported in several experiments (see \cite{FalconARFM2022} for a review), and it depends on several factors such as wave steepness \cite{Denissenko2007,Falcon2007,NazarenkoJFM2010,DeikeJFM2015,AubourgPRF2017}, %Lukaschuk et al. 2009,,Cobelli et al. 2011,
bound waves~\cite{CampagnePRF2018,MichelPRF2018}, dissipation~\cite{DeikeJFM2015,AubourgPRF2016}, and the container's shape \cite{IssenmannPRE2013}, but not on the confinement as observed here and in~\cite{HassainiPRF2018}. Note that a steepening of the gravity-wave spectrum due to finite-size effects has been only numerically reported~\cite{ZhangJFM2022}.  

\begin{figure*}[b!]
\includegraphics[width=8cm]{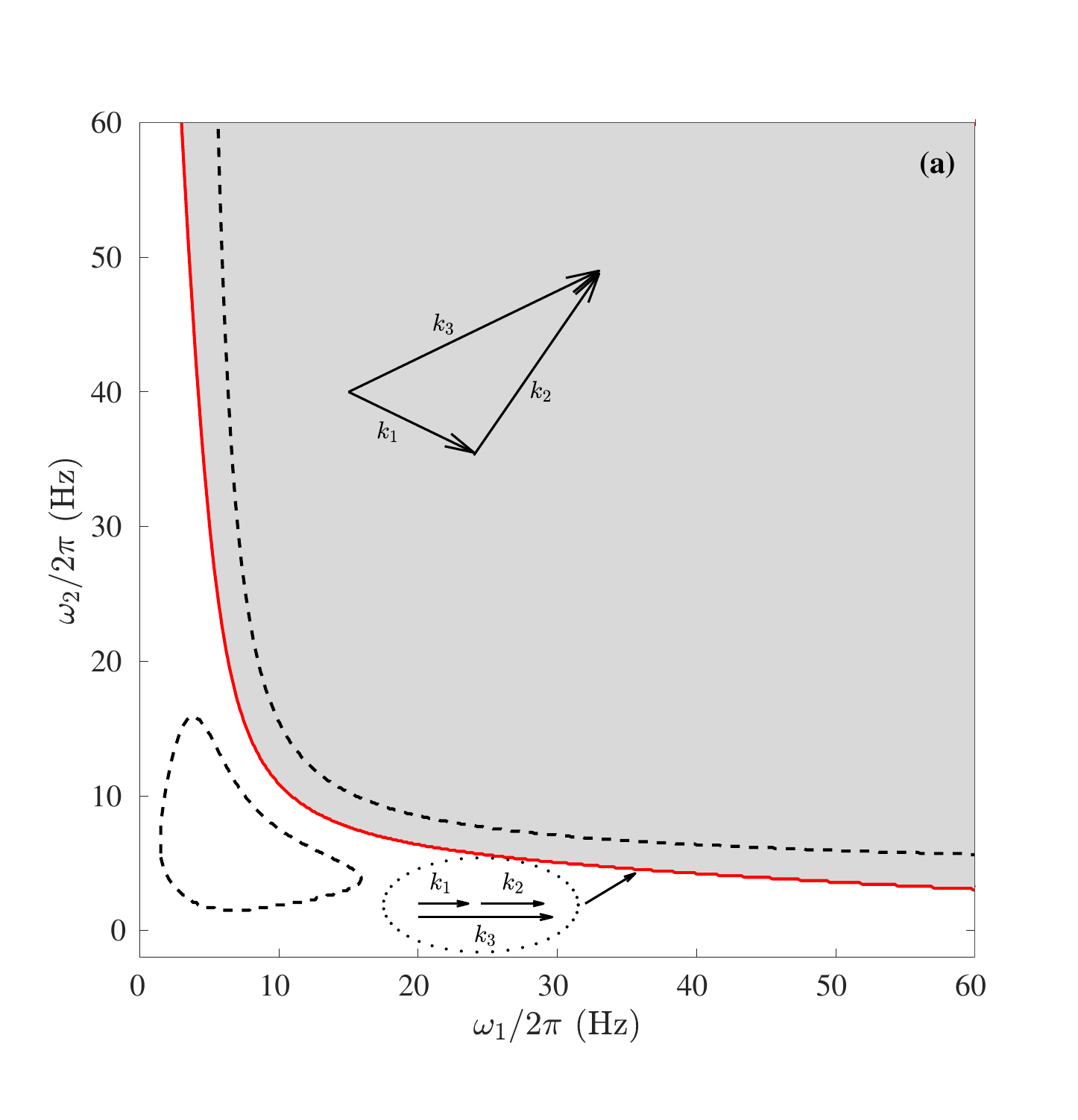}%a.eps} %[width=6.5cm,height=7.5cm] ,height=6.5cm
\includegraphics[width=8.5cm]{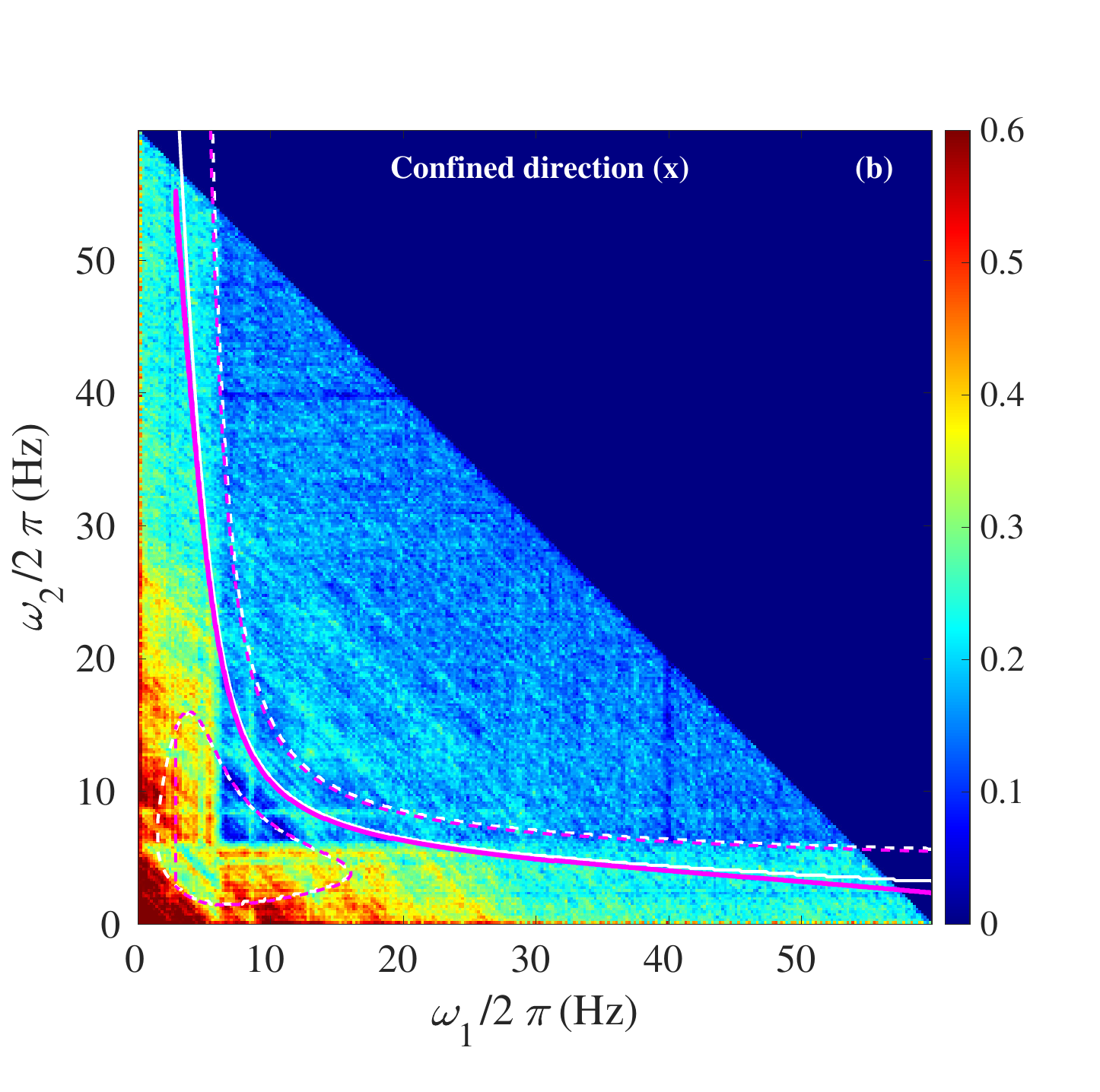}%7b.eps}

\includegraphics[width=8.5cm]{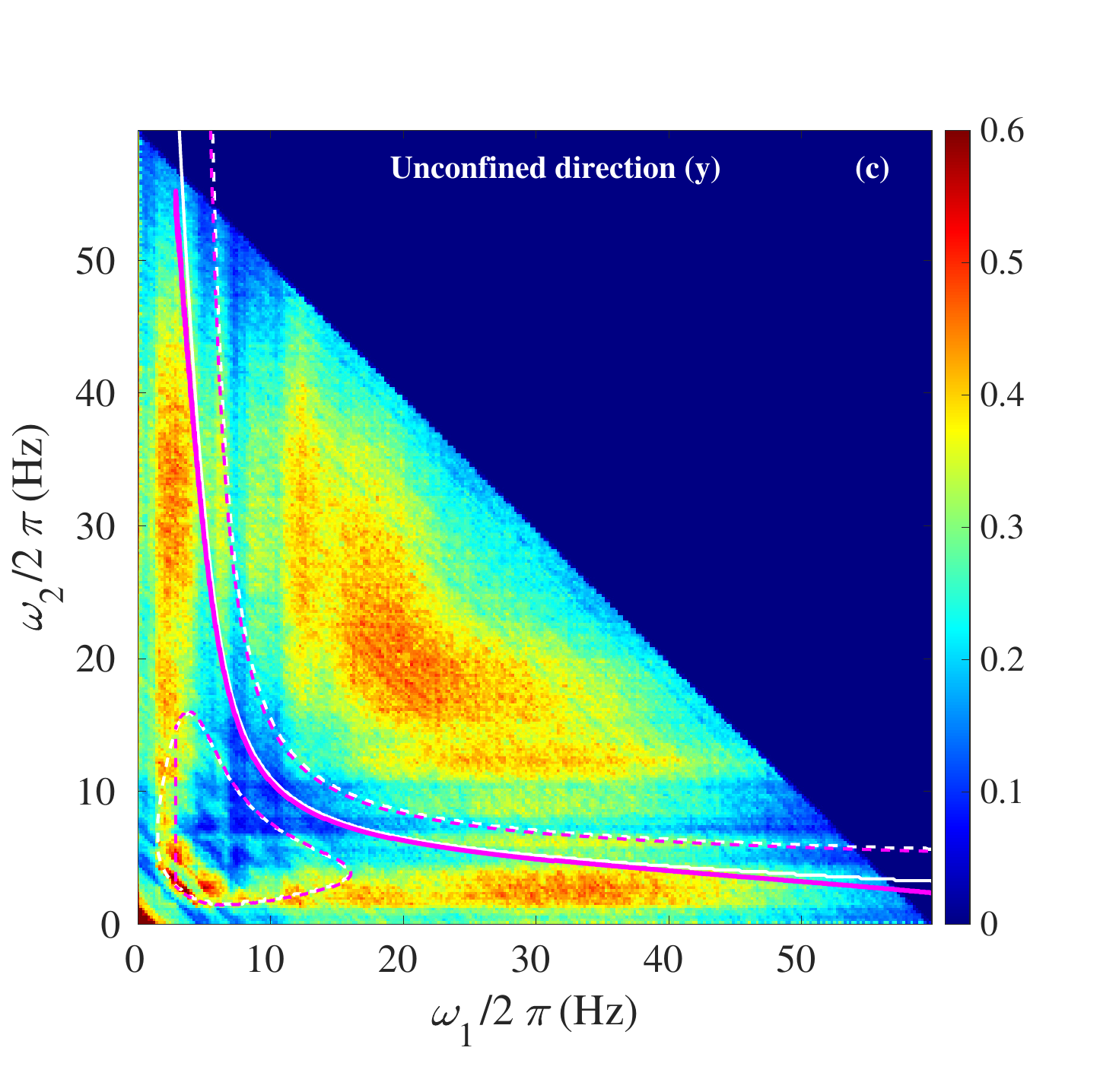}%7c.eps} 
\includegraphics[width=8.5cm]{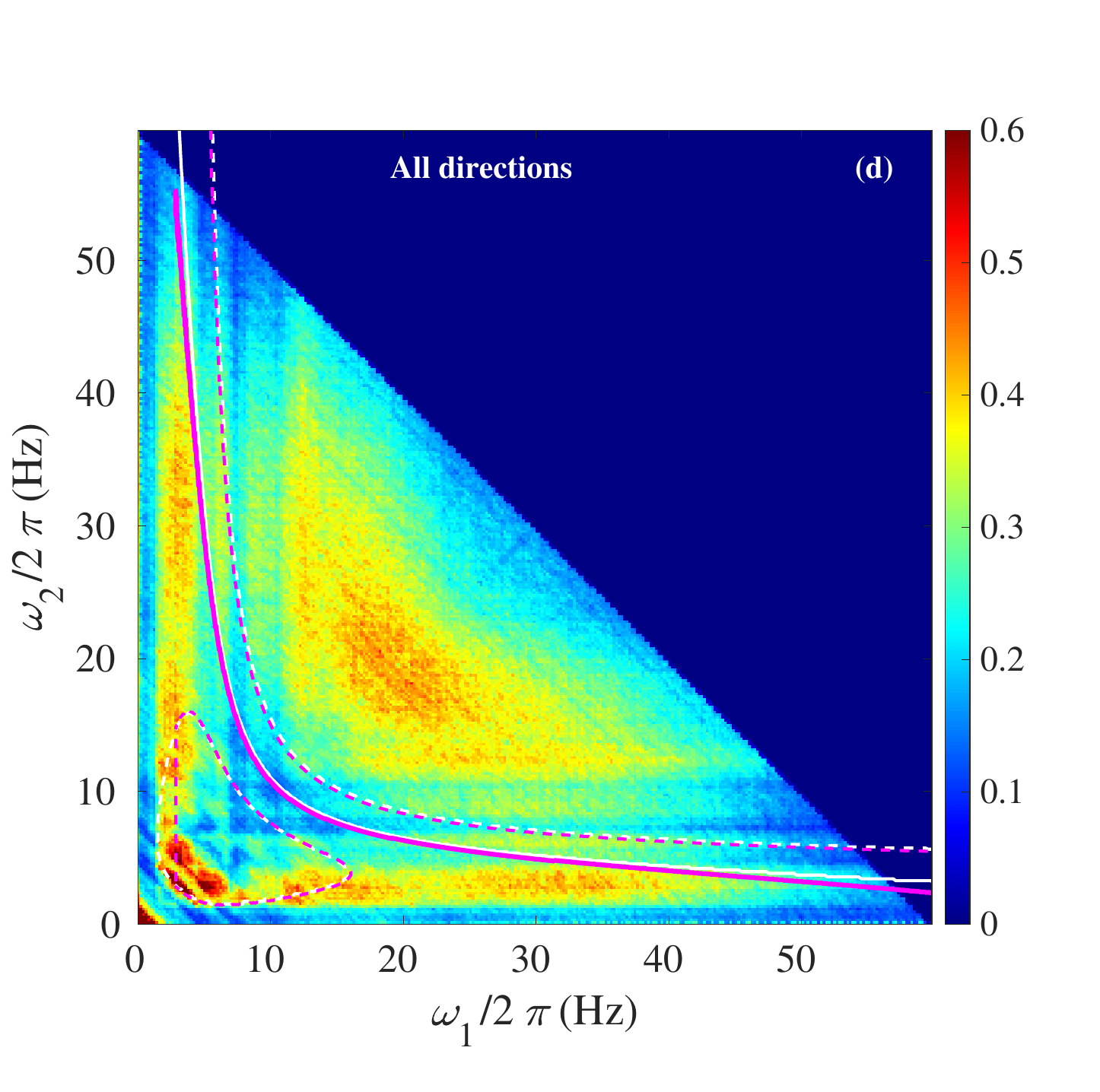}%7d.eps} ,height=7cm
\caption{(a) Theoretical solutions of three-wave resonant conditions of Eq.~\eqref{eq2} with $N=3$ and $\omega(k)$ given by Eq.~\eqref{eq1}. Red solid line marks the boundary case of 1D resonant interactions (i.e., three collinear waves), the gray region above this line indicates the existence of three-wave resonant interactions propagating in distinct directions. The region bounded by dashed lines, around the solid line, corresponds to quasiresonant 1D wave interaction conditions satisfying Eq.~\eqref{eq3} with scalar $k$, $N=3$, $\delta k/(2\pi) = 5$~m$^{-1}$ and $\omega(k)$ given by Eq.~\eqref{eq1}. (b-d) Experimental bicoherence [estimated from Eq.~\eqref{eq8}] of the wave vertical-velocity field showing the existence of three-wave interactions along  (b) the confined $x$-direction, (c) the unconfined $y$-direction, and (d) all directions (total wave field). Strong confinement $L_x=7$~cm (AR~$\simeq 7.1$), random forcing ($2\pm 0.5$~Hz), and $\epsilon=0.5$\%. White curves: solutions of interaction conditions of Eq.~\eqref{eq2} (solid) or Eq.~\eqref{eq3} (dashed), for $N=3$, between waves belonging only to the dispersion relation of Eq.~\eqref{eq1}. Pink curves: solutions of interaction conditions between waves from the first sloshing branch [Eq.~\eqref{eq6} with $n=1$] and waves from the dispersion relation of Eq.~\eqref{eq1}. The pink curves almost cover the white ones.}
\label{fig7} 
\end{figure*}

\subsection{Three-wave interactions}
Weakly nonlinear wave interactions lead to an energy transfer between waves towards smaller scales (as observed in Figs.~\ref{fig2} and~\ref{fig4}). At the lowest order in nonlinearity, three-wave interactions ($N=3$) are predicted to occur for pure capillary waves and for gravity-capillary waves to satisfy the resonance conditions of Eq.~\eqref{eq2}~\cite{NazarenkoBook,Zakharov,FalconARFM2022}. In contrast, four-wave interactions ($N=4$) are solutions for pure gravity waves. Let us consider three-wave interactions between gravity-capillary waves satisfying the linear dispersion relation of Eq.~\eqref{eq1}. The geometrical solution of Eq.~\eqref{eq2} with $N=3$ [i.e., $\omega_1+\omega_2=\omega_3$ and $\mathbf{k}_1+\mathbf{k}_2=\mathbf{k}_3$ with $\omega(k_i)$ given by Eq.~\eqref{eq1}] is shown in Fig.~\ref{fig7}(a) and highlights the domain of existence of resonant waves. No three-wave resonant solution exists for small frequency values in the ($\omega_1$, $\omega_2$) space, as waves are pure gravity waves. The red solid curve represents the boundary case where only 1D resonant interactions exist, in which two collinear gravity-capillary waves interact to create a third wave propagating in the same direction. The gray region above this solid line (i.e., for sufficiently large values of $\omega_1$ and $\omega_2$) represents the solutions of resonant interactions between waves propagating in distinct directions (i.e., 2D- or noncollinear-resonant interactions). Furthermore, as mentioned in Sect.~\ref{intro}, quasiresonant wave interactions can exist if they satisfy the conditions given by Eq.~\eqref{eq3}, where the exact resonant conditions of Eq.~\eqref{eq2} have been relaxed by $\delta k$ that quantifies the nonlinear broadening of the dispersion relation. The region, where these quasiresonant interaction solutions exist, is bounded by the dashed lines in Fig.~\ref{fig7}(a) around the solid line.

To probe the possible existence of three-wave interactions in our experiments, we compute the normalized third-order correlations in frequency of the wave vertical-velocity field, or bicoherence~\cite{PunzmannPRL2009,AubourgPRF2016,RicardEPL2021}
\begin{equation}
B(\omega_1,\omega_2) \equiv \frac{|\langle \hat{v}_{x,y}(\omega_1)\hat{v}_{x,y}(\omega_2)\hat{v}^{\ast}_{x,y}(\omega_1+\omega_2)\rangle|}{\sqrt{\langle |\hat{v}_{x,y}(\omega_1)\hat{v}_{x,y}(\omega_2)|^2\rangle \langle |\hat{v}_{x,y}(\omega_1+\omega_2)|^2\rangle}}\ ,
\label{eq8}
\end{equation}
where $\hat{v}_{x,y}(\omega)=\int_{\mathcal{T}} v(x,y,t)e^{-i\omega t}\,dt$ is the frequency Fourier transform of wave vertical velocity at ($x,y$) over $\mathcal{T}=20$~min, $^\ast$ denotes its complex conjugate, and $\langle \cdot \rangle$ represents an average over the entire space. The normalization is chosen to bound $B(\omega_1,\omega_2)$ between 0 (no correlation) and 1 (perfect correlation). Figure~\ref{fig7}(b-d) then shows the experimental bicoherences for a strong confinement ($L_x=7$~cm) computed along the confined direction [Fig.~\ref{fig7}(b)], the unconfined direction [Fig.~\ref{fig7}(c)], and for all directions [Fig.~\ref{fig7}(d)]. We also plot the theoretical solutions of resonant interactions of Eq.~\eqref{eq2} (white solid lines) and of quasiresonant interactions of Eq.~\eqref{eq3} with $N=3$ and a nonlinear broadening of $\delta k/(2\pi)=5$ m$^{-1}$ (white dashed lines). Pink curves correspond to the solutions of interaction conditions between waves from the first sloshing branch [Eq.~\eqref{eq6} with $n=1$] and waves from the dispersion relation of Eq.~\eqref{eq1}. The pink curves almost cover white ones, corresponding to waves that belong only to the dispersion relation of Eq.~\eqref{eq1}.

In the confined direction [Fig.~\ref{fig7}(b)], very few wave interactions are observed. This is because modes are depleted in this direction due to confinement, and thus they take discrete values. The frequencies $\omega_1$ (or $\omega_2$) of the red region on the figure are below the forcing frequency ($\simeq 2$~Hz). Henceforth, they have no significance in terms of bicoherence. However, the green-cyan region, above the solid line of the 1D resonant interaction, shows 2D resonant interactions in the confined direction, filling almost all the ($\omega_1$, $\omega_2$) parameter space homogeneously. This result contrasts strongly with the dotted pattern observed when the confined container is horizontally oscillating~\cite{AubourgPRF2016} rather than homogeneously forced as here. The bicoherence along the unconfined direction [Fig.~\ref{fig7}(c)] shows much stronger 2D resonant interactions. No collinear (1D) resonant interaction is observed here, in contrast to the ones reported when the container is horizontally oscillating~\cite{AubourgPRF2016}, which dominate and mask the confinement effects reported here.  We see in Fig.~\ref{fig7}(b) that 2D resonant wave interactions along the confined direction are strongly depleted with respect to the ones in Fig.~\ref{fig7}(c) along the unconfined direction. Indeed, Fourier modes are not discrete in the unconfined direction, and more waves belonging to the dispersion relation and the sloshing branches [Eq.~\eqref{eq6}] are present to interact.  The bicoherence of the full wave field (i.e., in all directions) in Fig.~\ref{fig7}(d) resembles that of the unconfined direction, which is not affected by the confinement. Finally, we have verified that the bicoherence, in the unconfined case, yields results qualitatively similar to those in Fig.~\ref{fig7}(c,d).  %, as well as quasiresonant interactions

\subsection{Discreteness timescale}
\begin{figure}[h!]
    \centering
  \includegraphics[width=0.55\linewidth]{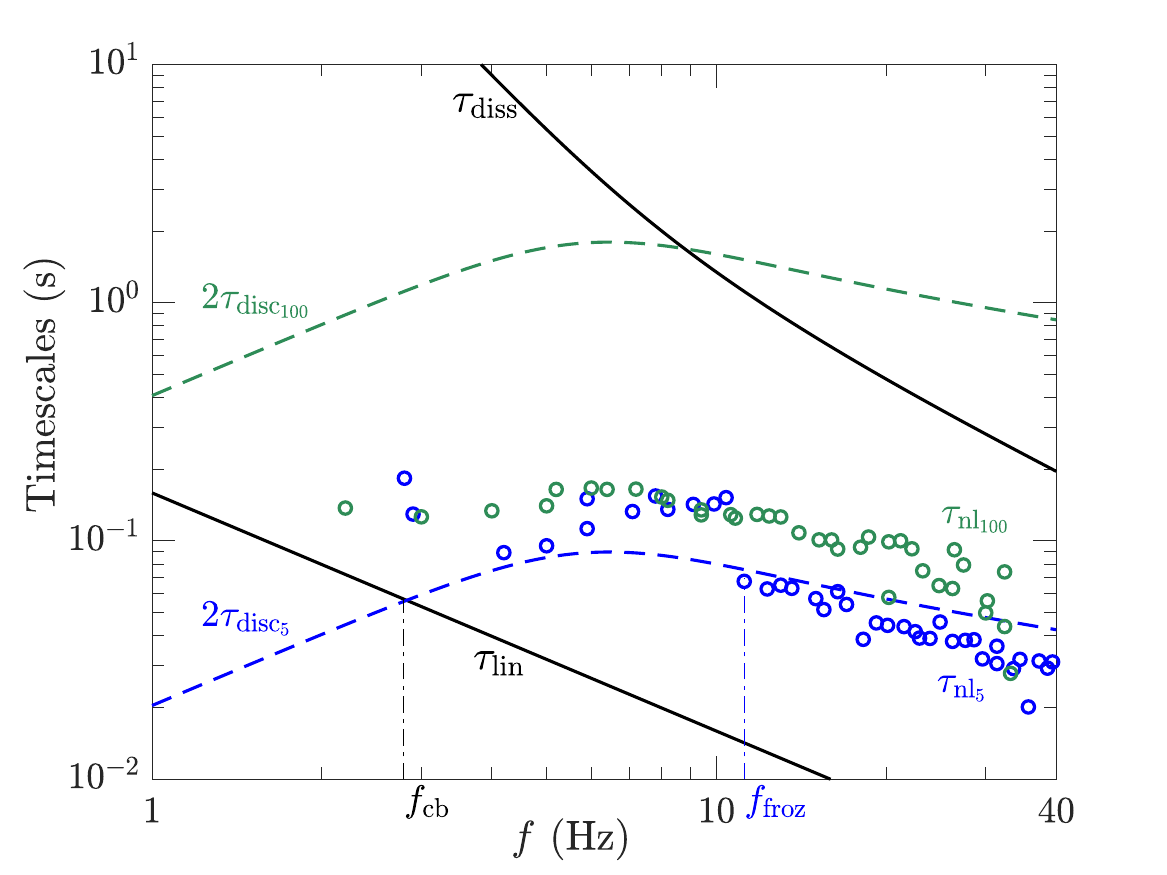} %timescales
    \caption{Wave turbulence timescales as a function of wave frequency. Bottom solid-black line: linear timescale $\tau_{\mathrm{lin}}=1/\omega$. Circles: experimental nonlinear timescale $\tau_{\mathrm{nl}}$ for the strongest confinement ($L_x=5$~cm - blue) and unconfined case ($L_x=100$~cm - green), estimated from Fig.~\ref{fig2}(a) and Fig.~\ref{fig2}(d). Top black curve: linear viscous dissipation timescale $\tau_{\mathrm{diss}}$ (see text). Dashed lines: discreteness time $\tau_{\mathrm{disc}}$ of Eq.~\eqref{disc} for the strongest confinement ($L_x=5$~cm - blue) and unconfined case ($L_x=100$~cm - green). Same colors as in Fig.~\ref{fig5}. The critical frequency $f_{\mathrm{froz}}$ separates discrete wave turbulence ($f<f_{\mathrm{froz}}$) from mesoscopic continuous wave turbulence ($f>f_{\mathrm{froz}}$). $f_{\mathrm{cb}}$ corresponds to a critical balance frequency.}  %s}
    \label{TimescaleFig}
\end{figure}

Weak turbulence theory assumes a timescale separation (regardless of $\omega$ in the inertial range)~\cite{NazarenkoBook}, between the linear time $\tau_{\mathrm{lin}}$, the nonlinear time $\tau_{\mathrm{nl}}$, the dissipation time $\tau_{\mathrm{diss}}$ (quantifying dissipative effects), and the discreteness time $\tau_{\mathrm{disc}}$ (quantifying finite-system size effects)~\cite{FalconARFM2022,RicardEPL2021,RicardPRF2023}, as
\begin{equation}
    \tau_{\mathrm{lin}}(\omega)\ll\tau_{\mathrm{nl}}(\omega)\ll[\tau_{\mathrm{diss}}(\omega); \tau_{\mathrm{disc}}(\omega)]. 
    \label{scale}
\end{equation}
The nonlinear evolution is thus assumed to be slow compared to the fast linear oscillations (wave period) but short compared to the typical wave dissipation time and the time linked to finite-size effects, enabling an energy cascade to occur in the inertial range. The evolutions of these timescales with frequency are plotted in Fig.~\ref{TimescaleFig}. The linear timescale is defined as $\tau_{\mathrm{lin}}=1/\omega$ (black solid line). The nonlinear timescale $\tau_{\mathrm{nl}}$ (circles) is experimentally inferred from the wave energy broadening around the dispersion relation of Fig.~\ref{fig2}, as $1/\delta\omega$ with $\delta\omega$ the full-width-at-half maximum of a Gaussian fit at each wavenumber. The dissipation timescale $\tau_{\mathrm{diss}}$ (black solid line) is computed as $\tau_{\mathrm{diss}}=2\sqrt{2}/[k(\omega)\sqrt{\nu\omega}]$, the main viscous contribution from the surface boundary layer with an inextensible film~\cite{Lamb1932,DeikePRE2014}, as contaminants are present~\cite{NoteDiss}.  The discreteness time $\tau_{\mathrm{disc}}$ (dashed lines) is computed as $\tau_{\mathrm{disc}}=1/\Delta \omega_{\mathrm{disc}}$ with $\Delta \omega_{\mathrm{disc}}=(\partial\omega/\partial k)\Delta k$~\cite{FalconARFM2022} and $\Delta k =2\pi/L_x$, the first eigenmode of the tank, that is %the first eigenmode in the confined direction
\begin{equation}
    \tau_{\mathrm{disc}}(\omega)=\left(\frac{\partial\omega}{\partial k}\Delta k\right)^{-1}=\frac{L_x}{\pi}\sqrt{\frac{k}{g}}\frac{\sqrt{1+(k/k_{gc})^2}}{1+3(k/k_{gc})^2},
    \label{disc}
\end{equation}
where $k(\omega)$ is as in Eq.~\eqref{eq1}, and $k_{gc}\equiv\sqrt{\rho g/\gamma}$ is the crossover wavenumber between gravity and capillary regimes (the inverse of the capillary length). The discreteness time of Eq.~\eqref{disc} is maximum for $k_{gc}/3$, and reads $\tau^g_{\mathrm{disc}}(\omega)=\frac{L_x}{\pi}\sqrt{\frac{k}{g}}=\frac{L_x}{\pi}\frac{\omega}{g}$ for pure gravity waves and $\tau^c_{\mathrm{disc}}(\omega)=\frac{L_x}{3\pi}\sqrt{\frac{\rho}{\gamma k}}=\frac{L_x}{3\pi}(\frac{\rho}{g\omega})^{1/3}$ for capillary waves~\cite{FalconARFM2022}. No discreteness effect is expected for $\tau_{\mathrm{nl}}(\omega)<2\tau_{\mathrm{disc}}(\omega)$, i.e., when the nonlinear spectral widening is larger than the half-frequency separation between adjacent eigenmodes. When $\tau_{\mathrm{nl}}(\omega)>2\tau_{\mathrm{disc}}(\omega)$, discrete or mesoscopic turbulence should occur, that is for $f<f_{\mathrm{froz}}$. The critical (or ``frozen") frequency $f_{\mathrm{froz}}$ is defined as $\tau_{\mathrm{nl}}(f_{\mathrm{froz}})=2\tau_{\mathrm{disc}}(f_{\mathrm{froz}})$ (see Fig.~\ref{TimescaleFig}). These finite-size effects are highlighted in the spectra along the confinement direction of Fig.~\ref{fig5}a by the emergence of a well-defined series of local peaks. When the confinement is gradually relaxed, a smooth transition towards a continuum spectrum is observed in Fig.~\ref{fig5}a, in agreement with the timescales of Fig.~\ref{TimescaleFig}. Indeed, the timescale-separation assumption of wave turbulence, as given by Eq.~\eqref{scale}, is well validated experimentally in the inertial range for the unconfined case ($L_x=100$~cm), thus confirming a wave turbulence spectrum up to 20~Hz. As the discreteness time decreases with decreasing $L_x$, finite-size effects become more significant, thereby reducing the inertial range of a continuous spectrum in favor of a discrete spectrum.
Finally, note that another critical balance occur when $\tau_l(\omega_{\mathrm{cb}}) = \tau_{\mathrm{disc}}(\omega_{\mathrm{cb}})$, that is for $f_{\mathrm{cb}}\simeq \sqrt{g/(8\pi L_x)}$, strongly breaking the timescale separation hypothesis for $f<f_{\mathrm{cb}}$. This balance is not achieved experimentally, as it occurs within the forcing range ($2\pm 0.5$~Hz).  
%see  \cite{FalconARFM2022,RicardEPL2021} and These Ricard p.31 and Annex A and \cite{RicardPRF2023} p.11-12)}
%\vspace{-0.5cm}

\section{Conclusions}
We experimentally investigated the influence of the finite-container size on weakly nonlinear random gravity-capillary surface waves. By locally forcing the fluid with magnets driven erratically by electromagnetic coils, we generated a homogeneous and isotropic random wave field without the dominant effect of global forcing, such as in horizontally oscillated tanks. Our spatiotemporal measurements reveal multiple branches in the spatiotemporal wave-energy spectrum in the unconfined direction, corresponding to sloshing modes of the confined direction. We showed that their cutoff frequencies at zero wavenumber and spectral properties can be tuned by varying either the confinement or the wave steepness. Moreover, the frequency wave spectrum in the confined direction displays discrete peaks in contrast to the continuous frequency power law in the unconfined direction. Using high-order correlation analysis, we demonstrate that two-dimensional three-wave resonant interactions are significantly depleted in the confined direction, whereas the unconfined direction retains these resonant interactions, resulting in continuous wave turbulence. These results thus indicate discrete wave turbulence in the confined direction, and mesoscopic wave turbulence in the unconfined direction, as the continuous frequency spectrum in this direction is also affected by confinement. As the confinement is gradually relaxed, we further show a smooth transition from discrete to continuous wave turbulence, consistent with the respective behaviors of the nonlinear and discreteness timescales. These findings establish that finite-system size effects deeply alter wave turbulence by reshaping spectral distributions and depleting two-dimensional resonant interactions along confined directions. They thus open the way to a more systematic understanding of how geometry constrains wave turbulence, with implications in both laboratory and geophysical systems. %no collinear (1D) interaction is observed here

%This study shows the finite-size effects on surface waves in water using a novel forcing method, wherein small magnets were used to generate waves. In the absence of confinement in the system, this forcing method generates waves with homogeneous properties. The principal effect of the confinement is the discreteness of the modes that can exist, which in turn gives rise to further effects. The wavelengths of these modes can be fine-tuned by adjusting the system's confinement. Furthermore, because of the discrete modes, sloshing branches were observed in the unconfined direction, and by systematically varying the confinement of the container and the energy injected into the system, we showed how these branches evolve as the spatio-temporal spectra of surface waves change. From the spatio-temporal spectra, cutoff frequencies of the sloshing branches were obtained, and a scaling law could be identified to show the universal nature of the branches. The temporal spectra along the confined and unconfined directions were also obtained, and it was shown how they differed from each other. The analysis of wave interactions via bicoherence revealed a clear suppression of resonant interactions in the confined direction. In contrast, the unconfined direction preserved the conditions of wave interaction as predicted by KWT. These results indicate a transition from KWT to DWT in the confined direction, and signatures of MWT in the unconfined direction, as the spectra in this direction are also affected by confinement.}

\begin{acknowledgments}
We thank Y. Le Goas and A. Di Palma for technical support. This work was supported by the Simons Foundation MPS-WT-00651463 Project (U.S.) on Wave Turbulence and the French National Research Agency (ANR Sogood Project No.ANR-21-CE30-0061-04 and ANR Lascaturb Project No. ANR-23-CE30-0043-02).
\end{acknowledgments}

%A UTILISER AVEC BIBTEX
%%\bibliographystyle{apsrev4-2}
%\bibliography{biblio} % Produces the bibliography via BibTeX.

%COPIE DU FICHIER BBL GENERE PAR BIBTEX
\end{document}